\documentclass[a4paper,nocompress]{spie}

\usepackage{amsmath,amsfonts,amssymb}
\usepackage{bm}   
\usepackage{graphicx}
\usepackage{astro_bib_macro}
\usepackage[normalem]{ulem}
\usepackage{booktabs}
\usepackage{listings}
\usepackage{lineno}

\usepackage[colorlinks=true, allcolors=blue]{hyperref} 

\def\Var{{\textrm{Var}}\,}

\DeclareMathOperator{\tr}{tr}

\graphicspath{{./}{./figures/}}   

\title{Analytical tolerancing of segmented telescope co-phasing for exo-Earth high-contrast imaging}

\author{Iva Laginja\supit{a}\supit{b}\supit{c}, R\'{e}mi Soummer\supit{a}, Laurent M. Mugnier\supit{b}, Laurent Pueyo\supit{a}, Jean-Fran\c{c}ois Sauvage\supit{b}\supit{c}, Lucie Leboulleux\supit{d}, Laura Coyle\supit{e}, J. Scott Knight\supit{e}
\skiplinehalf
\supit{a} Space Telescope Science Institute, Baltimore, MD 21218, USA\\
\supit{b} DOTA, ONERA, Universit\'e Paris Saclay, F-92322 Ch\^{a}tillon, France\\
\supit{c} Aix Marseille Universit\'{e}, CNRS, LAM (Laboratoire d'Astrophysique de Marseille) UMR 7326, 13388 Marseille, France\\
\supit{d} LESIA, Observatoire de Paris, Universit\'{e} PSL, CNRS, 92195 Meudon, France\\
\supit{e} Ball Aerospace \& Technologies Corp., Boulder, CO 80301, USA
}

\authorinfo{Further author information, send correspondence to Iva Laginja: E-mail: iva.laginja@lam.fr}

\begin{document} 
\maketitle

\begin{abstract}
This paper introduces an analytical method to calculate segment-level wavefront error tolerances in order to enable the detection of faint extra-solar planets using segmented-aperture telescopes in space. 
This study provides a full treatment of the case of spatially uncorrelated segment phasing errors for segmented telescope coronagraphy, which has so far only been approached using ad hoc Monte-Carlo simulations. Instead of describing the wavefront tolerance globally for all segments, our method produces spatially dependent requirement maps. We relate the statistical mean contrast in the coronagraph dark hole to the standard deviation of the wavefront error of each individual segment on the primary mirror. This statistical framework for segment-level tolerancing extends the Pair-based Analytical model for Segmented Telescope Imaging from Space (PASTIS), which is based uniquely on a matrix multiplication for the optical propagation. We confirm our analytical results with Monte-Carlo simulations of end-to-end optical propagations through a coronagraph. Comparing our results for the Apodized Pupil Lyot Coronagraph designs for the Large UltraViolet Optical InfraRed (LUVOIR) telescope to previous studies, we show general agreement but we provide a relaxation of the requirements for a significant subset of segments in the pupil. These requirement maps are unique to any given telescope geometry and coronagraph design. The spatially uncorrelated segment tolerances we calculate are a key element of a complete error budget that will also need to include allocations for correlated segment contributions. We discuss how the PASTIS formalism can be extended to the spatially correlated case by deriving the statistical mean contrast and its variance for a non-diagonal aberration covariance matrix. The PASTIS tolerancing framework therefore brings a new capability that is necessary for the global tolerancing of future segmented space observatories. 
\end{abstract}


\section{INTRODUCTION}
\label{sec:introduction}

Imaging Earth-like exoplanets and searching for biomarkers is one of the key science objectives in space astronomy for the next decade. The close proximity of such planets to their host star, as well as a flux ratio on the order of $10^{-10}$ at visible wavelengths makes this a challenging task. These two goals can be achieved by using large-aperture telescopes for large light collecting areas and high angular resolution, in combination with static and dynamic starlight suppression techniques with coronagraphs and wavefront sensing and control (WFS\&C) methods.

The invention of the coronagraph\cite{Lyot1939} synthesized with early ideas for the direct imaging of planets\cite{Roman1959} have led to several space mission concepts being developed toward this goal today. The Habitable Exoplanet Observatory\cite{Gaudi2020} (HabEx) and the Large UV Optical InfraRed Surveyor\cite{TheLUVOIRTeam2019, Bolcar2019} (LUVOIR) are two space-based concepts recently studied by NASA as possible future flagship missions. Their primary science objective is the direct detection and spectral characterization of habitable Earth-like planets\cite{Roberge2019} and the search for life; they require primary mirror diameters of 4--15 meters. Meanwhile, the ground-based community is preparing for the era of extremely large telescopes (ELTs) where 30--40 meter class telescopes like the Thirty Meter Telescope\cite{Simard2016} (TMT), the Giant Magellan Telescope\cite{Fanson2018} (GMT) and the European Extremely Large Telescope\cite{Ramsay2020} (E-ELT) will be equipped with coronagraphs and extreme adaptive optics systems to search for and characterize exoplanets\cite{Guyon2012}.

What unites all of these observatories is that they have significantly larger primary mirrors than their respective space-based and ground-based predecessors. This poses a number of problems that need to be solved, including considerations about overall mass, cost, and plausible launch vehicles for space-based missions. The logical consequence to this is that most of these observatories will have segmented primary mirrors, much like the Keck telescope\cite{Mast1982} or the James Webb Space Telescope (JWST)\cite{Acton2012, Perrin2018}. This will allow for lighter-weight backplanes and foldable structures for launch purposes, or even in-situ space assembly\cite{Polidan2018, Bowman2018}.

Telescope segmentation introduces additional diffraction effects in the focal plane\cite{Lightsey2003, Yaitskova2003, Troy2003, Itoh2019}, as well as sources for wavefront errors (WFE) due to segment misalignments and lighter mirror structure deformation in the form of localized segment-level aberration modes. All wavefront errors degrade the imaging performance in a high-contrast system\cite{Crossfield2007, Yaitskova2011} as they generate light residuals all over the focal plane. Such WFE will directly impact the performance of the coronagraph instrument. There are a number of coronagraph designs that were developed specifically to maximize performance on telescopes with arbitrary apertures, which includes secondary obscurations, spiders and segmentation gaps\cite{Sivaramakrishnan2005, Martinez2008, Soummer2009, Guyon2014, Zimmerman2016, Ruane2017}.

All high-contrast instruments that aim at very high-contrast such as what is necessary to detect Earth-like planets will deploy strategies that combine static coronagraph masks in pupil and focal planes with active control of the electric field\cite{Borde2006, Pueyo2009, Mazoyer2018a, Mazoyer2018b} in order to create a zone of deep contrast in the final image plane, the dark hole (DH). To enable such wavefront control techniques, several methods for focal plane wavefront sensing have been developed\cite{Groff2016, Jovanovic2018} to feed into a whole system of sensors and control loops that constitute the high-contrast instrument.

Even after careful cophasing of the segmented aperture and implementation of WFS\&C techniques that reach the required star attenuation level, there will always be some residual errors due to drifts in the system (e.g. from thermal instabilities). These changes to the mechanical structure and in the optical train will have a direct effect on the observability of a faint point source, as a sufficient signal-to-noise ratio is needed for detection within confidence limits\cite{Lyon2012}. As a consequence, these high-contrast goals with segmented apertures impose severe requirements not only on static wavefront quality, but also stability requirements on the WFE as well as the overall mechanical structures of the telescope. There are various works that have tried to quantify these wavefront stability requirements for high-contrast imaging, both with and without segmented apertures in mind. The Nancy Grace Roman Space Telescope (formerly known as the Wide-Field Infra-Red Survey Telescope, WFIRST) is a 2.4 meter monolithic space telescope with a large central obscuration and six thick, non-radial support struts\cite{Krist2015} that render high-contrast imaging particularly challenging\cite{Nemati2017wfirst}. Bound to launch in 2025, it will provide technology demonstrations for stellar coronagraphy at $10^{-9}$ contrast levels\cite{Savransky2016} with the Roman Space Telescope Coronagraphic Instrument (CGI). Going to a segmented telescope introduces an increased number of degrees of freedom that will influence the final contrast. While there are solutions that aim to maintain a good contrast in the dark hole across integration times by means of continuous WFS\&C\cite{Pogorelyuk2019}, the problem must also be approached from an overall engineering perspective\cite{Feinberg2017, Stahl2017, East2018}. In particular, the direct effects of segmentation on the final coronagraphic contrast\cite{Yaitskova2002, Stahl2015} are of interest in the context of high-contrast imaging, and there is an ongoing effort to characterize and quantify the requirements for such ultra-stable telescopes\cite{Coyle2018, Coyle2019, Pueyo2019, Hallibert2019}. Studies performing Monte-Carlo (MC) end-to-end (E2E) simulations\cite{Moore2018, Juanola-Parramon2019} have confirmed the strict WFE requirements of a couple of tens of picometers over tens of minutes to enable the search for faint extra-solar planets, and analytical methods for the derivation of coronagraphic performance specifications have been proposed\cite{Nemati2017, Nemati2020}.

One thing that all of these studies have in common is that they define global WFE tolerances over the entire telescope pupil, where the segments have a random contribution to the overall aberrations. In this paper, we focus on analytically defining requirements on a segment-to-segment basis instead, using the Pair-based Analytical model for Segmented Telescope Imaging from Space (PASTIS),\cite{Leboulleux2017, Leboulleux2018spie, Leboulleux2018jatis} which models the dark hole average contrast of a coronagraph on a segmented telescope as a function of the segment aberrations. We first introduce a new semi-analytical (SA) calculation method for the PASTIS matrix\cite{Laginja2019}. Then we show how to compute the statistical mean of the contrast using the PASTIS modes and extend the model inversion to calculate segment-level WFE requirements for a given target contrast.

A full error budget that aims at maintaining a particular DH contrast will contain WFE contributions both from spatially correlated as well as uncorrelated segments in the telescope pupil. The impact of aberrations made of correlated segments on coronagraph contrast has been studied in various cases, e.g. low-order Zernike modes as well as high-frequency checkerboard-like patterns in the pupil~\cite{N'Diaye2016, Moore2018, Douglas2018}. Aberrations made of spatially uncorrelated segments on the other hand have so far mostly been addressed in end-to-end simulations where the segments' amplitudes had equal standard deviations~\cite{Juanola-Parramon2019, Stahl2015, Nemati2017}. In this paper we tackle the uncorrelated contribution, and establish analytically how to allocate WFE contributions to all segments individually. This addresses an essential component in the overall error budget, which had not been formally established yet. We then use the PASTIS approach to also generalize this to the correlated case. 

\begin{figure}
   \begin{center}
   \begin{tabular}{c}
   \includegraphics[width = 12cm]{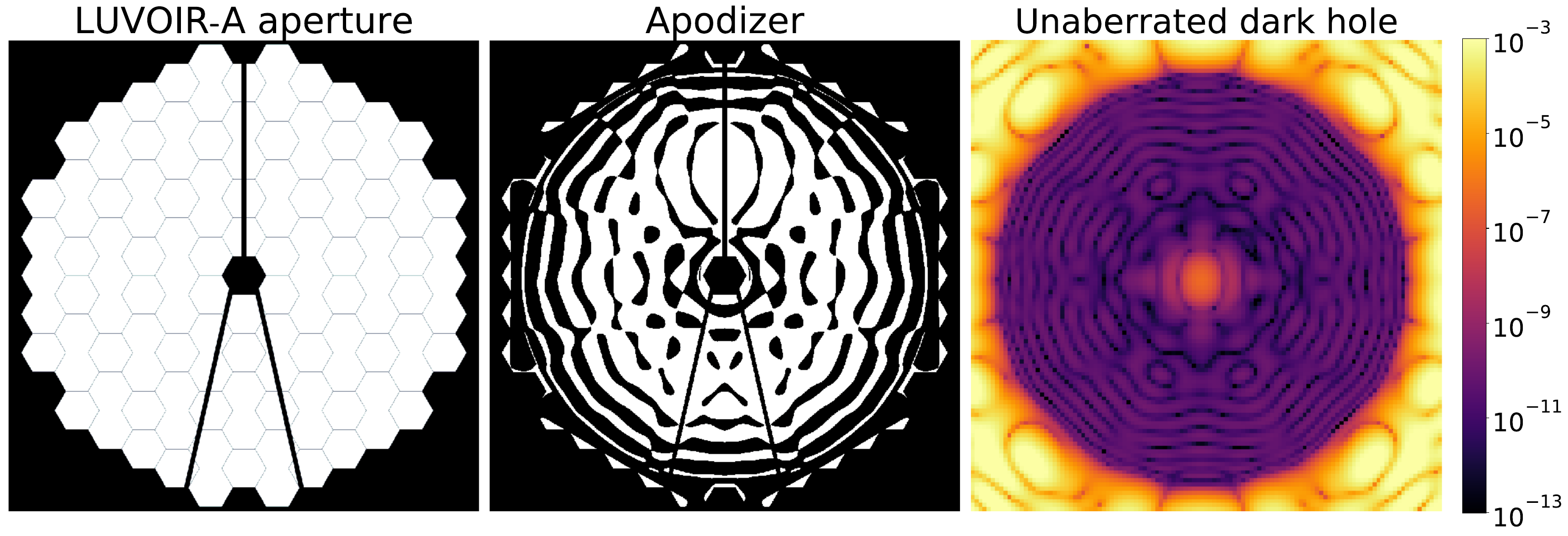}
   \end{tabular}
   \end{center}
   \caption[Aperture apod dh] 
   { \label{fig:aperture-apod-dh} 
    \textit{Left:} LUVOIR-A design aperture with a diameter of 15m. \textit{Middle:} Narrow-angle apodizer for the LUVOIR-A APLC, intended for exoplanet characterization. It uses a focal plane mask (FPM) with a radius of 3.5 $\lambda/D$ (with $\lambda$ the wavelength and $D$ the telescope diameter). \textit{Right:} Resulting coronagraphic image, with a dark hole from 3.4 to 12 $\lambda/D$ and an average normalized intensity of $4.3 \times 10^{-11}$, which is the coronagraph floor in the absence of optical aberrations.}
   \end{figure}

In Sec.~\ref{sec:top-level-pastis} we recall the development of the analytical propagation model and how the underlying PASTIS matrix was initially built through an analytical calculation. We then generalize the matrix calculation to all coronagraphs as well as segmented apertures with an extension to the semi-analytical matrix calculation, which eliminates the post-calibration step that used to be performed on a perfect coronagraph model. We show that the average contrast is always a quadratic function of the aberrations and drop the requirement of having a symmetrical dark hole. In Sec.~\ref{sec:model-inversion} we perform the model inversion and validate the semi-analytical matrix. Further, we show that the PASTIS modes can be used to define a statistical framework for the analysis, additionally to their deterministic relation to the dark hole contrast. In Sec.~\ref{sec:segment-error-budget} we derive the statistical mean contrast and its variance from two separate components that describe the imaging properties of the coronagraphic instrument on the one hand, and the thermo-mechanical segment statistical correlations on the other hand. We  proceed with the calculation of independent segment-based WFE requirements and how to validate them in a statistical sense, and we show how to apply this to correlated segments. All simulations in Sec.~\ref{sec:top-level-pastis}-\ref{sec:segment-error-budget} are done with a narrow-angle Apodized Pupil Lyot Coronagraph (APLC) on the primary pupil of LUVOIR-A (see Fig.~\ref{fig:aperture-apod-dh} and Sec.~\ref{sec:APPLICATION-TO-LUVOIR}) at a wavelength of 500~nm, however, these methods can be applied to any combination of coronagraph and segmented telescope. A full demonstration of the PASTIS analysis is given in Sec.~\ref{sec:APPLICATION-TO-LUVOIR}, where we calculate these segment tolerances for the case of three different APLC designs for LUVOIR-A and highlight some consequences of this approach. In Sec.~\ref{sec:DISCUSSION} we discuss our results and compare them to previously derived requirements and how they can be used in observatory error-budgets, and in Sec.~\ref{sec:CONCLUSION} we report our conclusions.

Note that the main metric of the PASTIS model is the spatial average raw contrast in the dark hole (normalized coronagraphic intensity to peak of direct image), which is what we refer to as ``contrast" throughout this paper, as opposed to a spatially dependent quantity. We also want to point out how we differentiate between this spatially averaged dark hole intensity, the ``average DH contrast" on the one side, and a statistical mean (expectation value) of this averaged contrast over many optical propagations on the other side, the statistical ``mean contrast".


\section{PASTIS model of telescope segment-level aberrations in high-contrast coronagraphy}
\label{sec:top-level-pastis}

The PASTIS model was initially established for a perfect coronagraph using an analytical propagation model for aberrated pairs of segments\cite{Leboulleux2018jatis}; the application to real coronagraphs required a second-step numerical calibration.  Here, we generalize the model to any coronagraph on any segmented aperture geometry by using a semi-analytical derivation of the PASTIS matrix. We also show that the validity of the PASTIS results is not limited to symmetrical dark holes, but extends to non-symmetrical ones as well. Independently of the way the PASTIS matrix is calculated (analytically or semi-analytically), the derivations and conclusions that we build on the PASTIS approach retain their analytical power and potential. 

\begin{figure*}[ht]
   \begin{center}
   \begin{tabular}{@{}c@{}}
   \includegraphics[width =\textwidth]{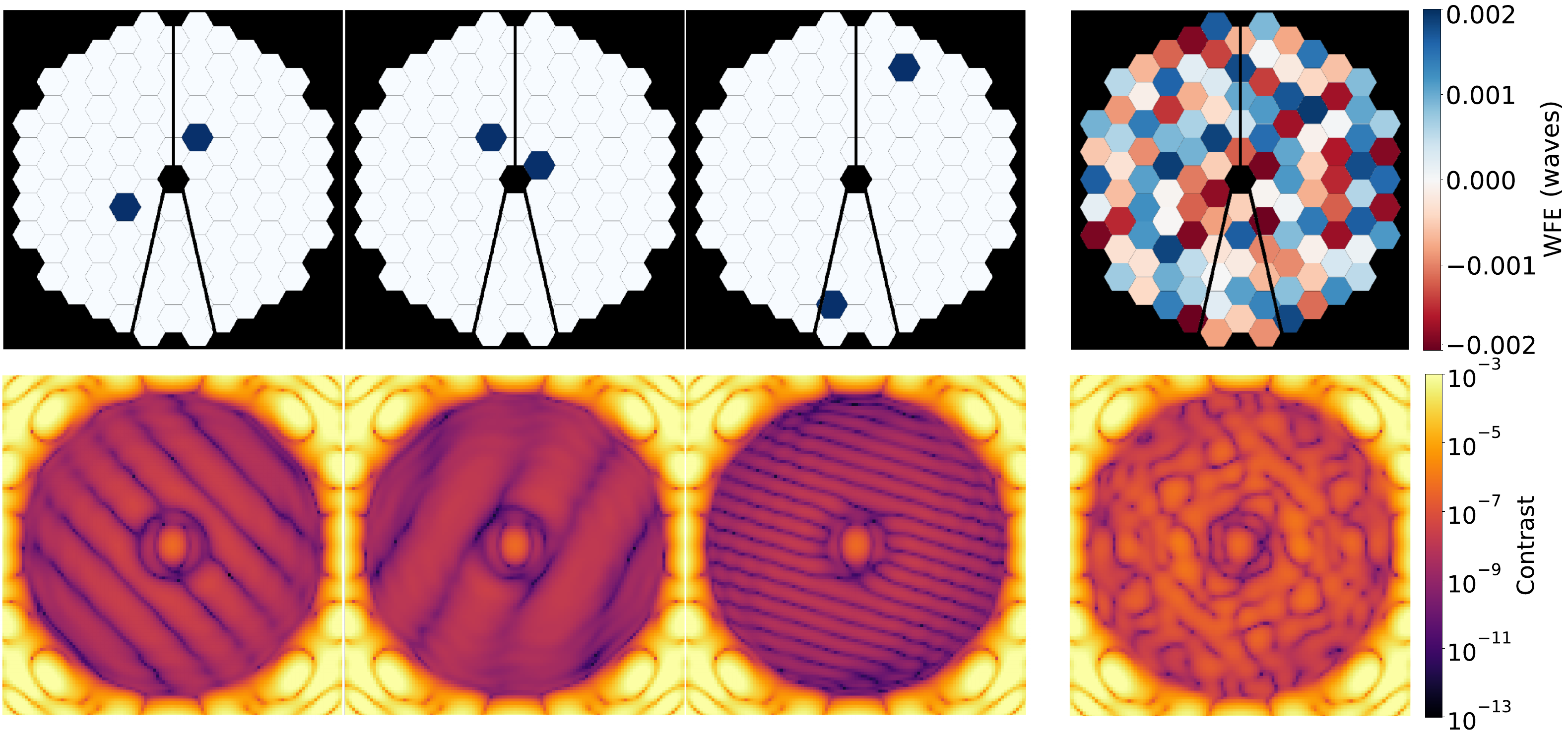}
   \end{tabular}
   \end{center}
   \caption[Piston Pairs] 
   { \label{fig:piston-pairs} 
    Piston pair aberrations on a segmented pupil (top) and the resulting image plane intensity distributions in the dark hole (bottom), using the narrow-angle APLC for LUVOIR-A in an E2E propagation model. The left three panels show different interference pairs with corresponding Young-like interference fringes, while the right panel shows a random distribution of local piston on all segments of the pupil and the resulting image plane intensity. All plots appear on the same scale.}
    \end{figure*}

\subsection{Matrix formalism to calculate the average dark hole contrast}
\label{subsec:pastis-introduction}

The goal of PASTIS is to model coronagraphic images in the presence of optical aberrations on a segmented primary mirror, which can be represented for example by using localized Zernike polynomials. This basis is an obvious possible choice since segment-level piston, tip/tilt, focus and astigmatism are naturally occurring aberrations from segment misalignments, for example in three-mirror anastigmat (TMA) designs such as JWST\cite{Acton2004, Acton2012} or LUVOIR\cite{TheLUVOIRTeam2019, Juanola-Parramon2019}. Although beyond the scope of this paper, the PASTIS approach can also be applied directly to any other function basis, for example to represent mirror wavefront errors induced by thermo-mechanical effects (mounting, backplane deformations, etc.)\cite{ULTRA-Phase1-report2019, East2019, Wells2019}. In this paper, we simply expand the phase aberration in the segmented pupil $\phi_s$ as a sum of local (segment-level) Zernike polynomials\cite[Eq.~9]{Leboulleux2018jatis}:
    \begin{equation}
    \phi_s (\mathbf{r}) = \sum_{(k,l)=(1,0)}^{(n_{seg}, n_{zer})} a_{k,l}\ Z_l(\boldsymbol{\mathbf{r}}-\mathbf{r_k}),
    \label{eq:local-aberrations}
    \end{equation}
where $\mathbf{r}$ is the pupil plane coordinate, $\phi_s$ the phase from the segmented primary and $n_{seg}$ is the total number of segments, indexed by $k$. The $a_{k,l}$ are the Zernike coefficients with Noll index\cite{Noll1976} $l$ up to the maximum Zernike $n_{zer}$, and $Z_l$ is the $l^{th}$ Zernike. 
In this paper, we limit the study to a single Zernike mode (piston; index $l=0$) as illustrated in Fig.~\ref{fig:piston-pairs}. Hence we drop the $l$ index in all consecutive equations, but the PASTIS methodology is applicable to any Zernike mode, combination thereof, or other types of segment-level modes.
   
In high-contrast coronagraphy the best contrast is not typically obtained for the perfect aperture without any aberration, but more commonly in the presence of a wavefront control solution using deformable mirrors\cite{Pueyo2019, Mazoyer2018a}. Therefore, we are studying the response of the coronagraphic system to a perturbation around that solution. Defining $\phi_{DH}$ as the phase solution for best DH contrast and $\phi_s$ as the segmented perturbation, the phase can be divided into
    \begin{equation}
    \phi = \phi_{DH} + \phi_s.
    \label{eq:composite-phase}
    \end{equation}
High-contrast coronagraphy requires exquisite wavefront quality around the dark hole solution, and therefore we assume the small aberration regime for $\phi_s$, where the electric field $E(\mathbf{r})$ is well approximated as an affine function of the phase: 
$E(\mathbf{r}) = P(\mathbf{r})\,e^{i\phi(\mathbf{r})}\simeq P'(\mathbf{r}) + i\,\phi_s(\mathbf{r})$. The phase $\phi_s(\mathbf{r})$ is zero where the pupil aperture $P(\mathbf{r})$ is zero, and $P'(\mathbf{r})$ is a complex pupil that includes the wavefront solution to produce the static dark hole (with both phase and amplitude contributions, and including static errors). Note that the phase $\phi_{DH}$ is not necessarily small\cite{Mazoyer2018a, Mazoyer2018b}.

Using Fourier optics for a scalar description of the electric field and of its propagation, the coronagraph propagation can be represented by a linear operator $\mathcal{C}$. This is a valid assumption for Lyot-style coronagraphs, for example an APLC\cite{Soummer2003, N'Diaye2015, N'Diaye2016} such as the one illustrated in Fig.~\ref{fig:aperture-apod-dh} for the LUVOIR-A coronagraph design, or a vortex coronagraph\cite{Foo2005, Mawet2013}. High-order vortex designs would need a special treatment for the specific low-order modes (e.g. defocus) they reject perfectly\cite{Ruane2017}, but the tolerancing of such global modes is not the main purpose of PASTIS anyway. We can hence express the intensity distribution in the final image plane as
    \begin{equation}
    I (\mathbf{s}, \phi) = | \mathcal{C}\{P'\}(\mathbf{s}) + i\, \mathcal{C}\{\phi_s\}(\mathbf{s})|^2,
    \label{eq:real-coro}
    \end{equation}
with $\mathbf{s}$ the image plane coordinate. This intensity is therefore the sum of three terms\cite[Eq.~16]{Leboulleux2018jatis}: 
    \begin{equation}
    I(\mathbf{s}) = | \mathcal{C}\{P'\} |^2 + 2 \Re\{\mathcal{C}\{P'\} \mathcal{C}\{\phi_s\}^*\} + |\mathcal{C}\{\phi_s\}|^2.
    \label{eq:intensity-three-terms}
    \end{equation}
In most cases of interest, we will be working in a symmetrical DH. It can be shown that the spatial average of the linear cross-term in Eq.~\ref{eq:intensity-three-terms} over a symmetrical DH is zero\cite[Appendix A]{Leboulleux2018jatis}. This simplifies Eq.~\ref{eq:intensity-three-terms} to a quadratic function of the phase:
    \begin{equation}
    \langle I(\mathbf{s}) \rangle_{DH} = \langle | \mathcal{C}\{P'\} |^2 \rangle_{DH} + \langle |\mathcal{C}\{\phi_s\}|^2 \rangle_{DH}.
    \label{eq:intensity-two-terms}
    \end{equation}
     
The main metric used in this paper is the spatial average contrast over the extent of the dark hole, $\langle \dots \rangle_{DH}$, so by using $c_0' = \langle | \mathcal{C}\{P'\} |^2 \rangle_{DH}$, we can express the average dark hole intensity as:
    \begin{equation}
    \langle I(\mathbf{s})\rangle_{DH} = c_0' + \langle |\mathcal{C}\{\phi_s\}|^2\rangle_{DH}.
    \label{eq:avg_intensity_quadratic}
    \end{equation}
Using the expression for the phase decomposition from Eq.~\ref{eq:local-aberrations} in Eq.~\ref{eq:avg_intensity_quadratic}, we can derive the intensity as a function of all aberrated segment pair combinations: 
    \begin{equation}
    \langle I(\mathbf{s})\rangle_{DH} = c_0' + \langle |\mathcal{C}\Big\{ \sum_k^{n_{seg}} a_k Z(\mathbf{r} - \mathbf{r}_k) \Big\}|^2\rangle_{DH},
    \label{eq:square-modulus}
    \end{equation}
    and therefore:
    \begin{align}
    \langle I(\mathbf{s})\rangle_{DH} &= c_0' + \nonumber \\
                                    & \sum_i^{n_{seg}} \sum_j^{n_{seg}} a_i a_j \langle \mathcal{C}\{Z(\mathbf{r} - \mathbf{r}_i)\} \mathcal{C}\{Z(\mathbf{r} - \mathbf{r}_j)\}^*\rangle_{DH}.
    \label{eq:intensity-from-pair-aberrations}
    \end{align}
This double sum combines all pairs of segments where segment $i$ has an aberration amplitude $a_i$ of the localized phase aberration $Z(\mathbf{r} - \mathbf{r}_i)$.  
These cross-terms from each aberrated pair of segments are very similar to Young interference fringes, and this forms the basic idea behind the PASTIS model\cite{Leboulleux2018jatis}. The orientation and periodicity of these fringes depend on the separation and orientation of the according aberrated pair, as displayed in Fig.~\ref{fig:piston-pairs}.

It is important to note that the pair-wise model is not an ad-hoc idea to build the model by pairs. It derives from the fact that we expand the primary mirror phase on a discrete number of segments. Since we build a propagation model for the intensity, the ``pairs" simply appear in Eq.~\ref{eq:intensity-from-pair-aberrations} from all the cross-terms when calculating the square modulus of the electric field in Eq.~\ref{eq:square-modulus}.

Eq.~\ref{eq:intensity-from-pair-aberrations} can be readily re-written as a matrix multiplication:
    \begin{equation}
    c = c_0' + \mathbf{a}^T M \mathbf{a},
    \label{eq:pastis-equation}
    \end{equation}
where $c$ is the average contrast in the dark hole, $c_0'$ the coronagraph floor (i.e. the average contrast in the dark hole at best contrast with $\phi_{DH}$, in the absence of phase perturbations), $M$ is the PASTIS matrix with elements $m_{ij}$, $\mathbf{a}$ is the aberration vector of the local Zernike coefficients on all discrete $n_{seg}$ segments and $\mathbf{a}^T$ its transpose.  The elements of the PASTIS matrix $M$ in Eq.~\ref{eq:pastis-equation} therefore directly identify as:
    \begin{equation}
    \begin{aligned}
    m_{ij} &= \langle \mathcal{C}\{Z(\mathbf{r} - \mathbf{r}_i)\} \mathcal{C}\{Z(\mathbf{r} - \mathbf{r}_j)\}^*\rangle_{DH}. \\
    \label{eq:matrix-elements}
    \end{aligned}
    \end{equation}

While this derivation is always true in the most common case of a symmetrical DH, there are coronagraph designs that produce half-sided dark holes\cite{Por2020}. We can show that the quadratic dependency of the contrast on the phase perturbations remains true in this most general case.  We rewrite Eq.~\ref{eq:intensity-three-terms} in a similar matrix form as Eq.~\ref{eq:pastis-equation}, but preserving the linear term:
    \begin{equation}
    c = c_0' + \mathbf{v}^T \mathbf{a} + \mathbf{a}^T M \mathbf{a},
    \label{eq:vectorized-linear}
    \end{equation}
where $\mathbf{v}$ is a vector that does not need to be expressed explicitly here. 
If we take the derivative of this equation and solve for the aberration vector $\mathbf{a_0}$ that provides the minimum contrast $c_0$, we can identify $\mathbf{a_0} = -M^{-1} \mathbf{v}/2$ and $c_0 = c_0' - 1/4 \mathbf{v}^T M^{-1} \mathbf{v}$, and therefore eliminate the linear term by performing a simple change of variable:
    \begin{equation}
    c = c_0 + (\mathbf{a}-\mathbf{a_0})^T M (\mathbf{a}-\mathbf{a_0}).
    \label{eq:vectorized-shifted}
    \end{equation}
This quadratic expression is similar to Eq.~\ref{eq:pastis-equation}, but with a segmented mirror perturbation solution $\mathbf{a_0} \neq 0$ that improves contrast compared to the case without aberrations. 
As discussed above, we also assume a wavefront control solution with deformable mirrors to be included in the term $P'(\mathbf{r})$ (and hence $c_0$), with both amplitude and phase contributions. Therefore this guarantees that the best contrast in the presence of that wavefront control solution and DH is obtained for $\mathbf{a_0}=0$, which in turn means that any arbitrary segment aberration vector $\mathbf{a}$ will always degrade the contrast. Note that this does not preclude to have a non-zero static segmented correction included as part of the term $\phi_{DH}$. This is equally true in broadband light: When summing over wavelengths, the quadratic nature of Eq.~\ref{eq:vectorized-shifted} remains true, albeit with different coefficients $c_0$, $a_0$ and $M$.

We have shown that the average dark hole contrast is always a quadratic function of a segmented phase perturbation $\phi_s$, which can be discretized into a per-segment aberration amplitude vector $\mathbf{a}$, coefficients on a modal basis. We can calculate this average dark hole contrast for any aberration vector directly, using the PASTIS matrix expression (Eq.~\ref{eq:pastis-equation}). This is particularly interesting and efficient since it does not require end-to-end optical simulations, and only involves simple linear algebra. Furthermore, this analytical expression can be inverted to establish a segment-level wavefront error budget that meets a given level of contrast. This will be detailed in Sec.~\ref{sec:segment-error-budget}.

\subsection{Semi-analytical calculation of the PASTIS matrix}
\label{subsec:semi-analytic-extension}
The PASTIS matrix $M$ can be calculated using the original analytical approach for a perfect coronagraph, then calibrated numerically for a real coronagraph and to include pupil features (e.g. support structures) \cite[Eq.~20]{Leboulleux2018jatis}. This approach was validated against an end-to-end model for the 36-segment ATLAST telescope pupil with an APLC \cite[Fig.~7]{Leboulleux2018jatis} to within an error of 3\%.

Here, we introduce another way to calculate the PASTIS matrix using an end-to-end simulation\cite{Laginja2020pastis} of the average dark hole contrast for all individually aberrated segment pairs, from which we can identify semi-analytically the matrix elements in Eq.~\ref{eq:matrix-elements}. This presents the advantage of enabling a direct calculation of the matrix for any telescope geometry, any coronagraph, and any choice of segment-level aberrations (including fully numerical ones such as segment figures induced by thermo-mechanical effects).

The phase for each segment pair is expressed as a Zernike aberration:
    \begin{equation}
    \phi_{ij}(\mathbf{r}) = a_i Z(\mathbf{r} - \mathbf{r}_i) + a_j Z(\mathbf{r} - \mathbf{r}_j).
    \label{eq:pair-phase}
    \end{equation}
We denote by $c_{ij} = \left \langle I_{ij}(\mathbf{s)} \right \rangle_{DH}$ the average dark hole contrast, on the pair of segments $i,j$, $\phi_{ij}(\mathbf{r})$, that can be calculated numerically for a small wavefront aberration and compared to the quadratic expression of Eq.~\ref{eq:avg_intensity_quadratic} under the linear expansion of this phase term: 
    \begin{equation}
    \begin{aligned}
    c_{ij} &= c_0 + \langle | a_i \mathcal{C}\{Z(\mathbf{r} - \mathbf{r}_i)\} + a_j \mathcal{C}\{Z(\mathbf{r} - \mathbf{r}_j)\} |^2 \rangle_{DH} \\
    &=  c_0 + a_i^2 \langle | \mathcal{C}\{Z(\mathbf{r} - \mathbf{r}_i)\} |^2 \rangle_{DH} + a_j^2 \langle | \mathcal{C}\{Z(\mathbf{r} - \mathbf{r}_j)\} |^2 \rangle_{DH} \\
    &+ a_i a_j 2 \langle \mathcal{C}\{Z(\mathbf{r} - \mathbf{r}_i)\} \mathcal{C}\{Z(\mathbf{r} - \mathbf{r}_j)\}^* \rangle_{DH}.
    \label{eq:three-terms}
    \end{aligned}
    \end{equation}
The elements $m_{ij}$ of the PASTIS matrix $M$ (Eq.~\ref{eq:matrix-elements}) can then be identified directly in Eq.~\ref{eq:three-terms} as:
    \begin{equation}
     c_{ij} =  c_0 + a_i^2 m_{ii} + a_j^2 m_{jj} + 2 a_i a_j\ m_{ij},
     \label{eq:almost-there-off}
    \end{equation}
where the diagonal terms of the PASTIS matrix are
    \begin{equation}
     m_{ii} = \frac{c_{ii} - c_0}{a_i^2},
     \label{eq:diagonal-elements}
    \end{equation}
and the off-diagonal elements\footnote{This equation corrects a sign error in a previous conference proceeding.\cite[Eq.~12]{Laginja2019}}:
    \begin{equation}
     m_{ij} = \frac{c_{ij} + c_0 - c_{ii} - c_{jj}}{2 a_i a_j}.
     \label{eq:off-diagonal-elements}
    \end{equation}

For simplicity, we choose the same calibration aberration amplitude $a_c = a_i = a_j$  for both segments. Throughout our analytical development above, $a_i$ is in units of radians, as $\phi$ is a phase. Since the PASTIS matrix can be normalized to any units though, the units of the aberration amplitude $a_c$ can be chosen freely in the computation of Eqs.~\ref{eq:diagonal-elements} and \ref{eq:off-diagonal-elements}. The units of the PASTIS matrix are therefore in ``contrast per square of units of $a_c$'' (contrast having no physical dimension), which is consistent with Eq.~\ref{eq:pastis-equation}.  Note that in the presented case in Fig.~\ref{fig:numerical_matrix}, the units of the aberration amplitude $a_c$ is waves. 
The aberration amplitude $a_c$ has to be chosen such that the global pupil aberration it results in yields an average DH contrast higher than the contrast floor, but small enough to remain in the small phase aberrations linear regime. This will be discussed further in Sec.~\ref{subsec:validating-pastis}. 

The matrix is symmetric by definition since $c_{ij} = c_{ji}$. Off-diagonal elements $m_{ij}$ of the PASTIS matrix (Eq.~\ref{eq:off-diagonal-elements}) can be negative, which is not an issue since the only constraint for the matrix is to be positive semi-definite to ensure positive eigenvalues, since they correspond to each mode's contrast. This will be discussed in detail in Sec.~\ref{sec:model-inversion}.

We could potentially calculate the matrix elements $m_{ij}$ (Eq.~\ref{eq:matrix-elements}) directly by calculating those complex electric field quantities. Usually though, full end-to-end simulators that calculate the image plane intensity are readily available and necessary for multiple other reasons. This means that choosing to calculate the PASTIS matrix through image plane intensities makes it more flexible and portable to other simulators. More importantly, working with intensities allows us to measure an empirical PASTIS matrix without the estimation errors and computational overheads of using an electric field estimator, allowing this theory to be experimentally tested.

In summary, the PASTIS matrix is constructed in two steps: (1) Calculate aberrated images $I_{ij}$ and their corresponding dark hole average contrast $c_{ij}$ for each pair of aberrated segments $i,j$, and (2) use these contrast values to identify analytically the elements of the PASTIS matrix $M$ based on Eqs.~\ref{eq:diagonal-elements} and \ref{eq:off-diagonal-elements}. Here, the numerical calculation of these aberrated images for pairs of segments using an end-to-end simulator (see Appendix \ref{appendix-simulator}) replaces the analytical expression of Young fringes between pairs of segments\cite{Leboulleux2018jatis}. This approach provides more accuracy, flexibility and generality for use with any coronagraph and telescope geometry, since the analytical approach has to be calibrated using a numerical simulation anyway. 

\subsection{Validating the semi-analytical PASTIS matrix}
\label{subsec:validating-pastis}
\begin{figure*}[ht]
   \begin{center}
   \begin{tabular}{c}
   \includegraphics[width=\textwidth]{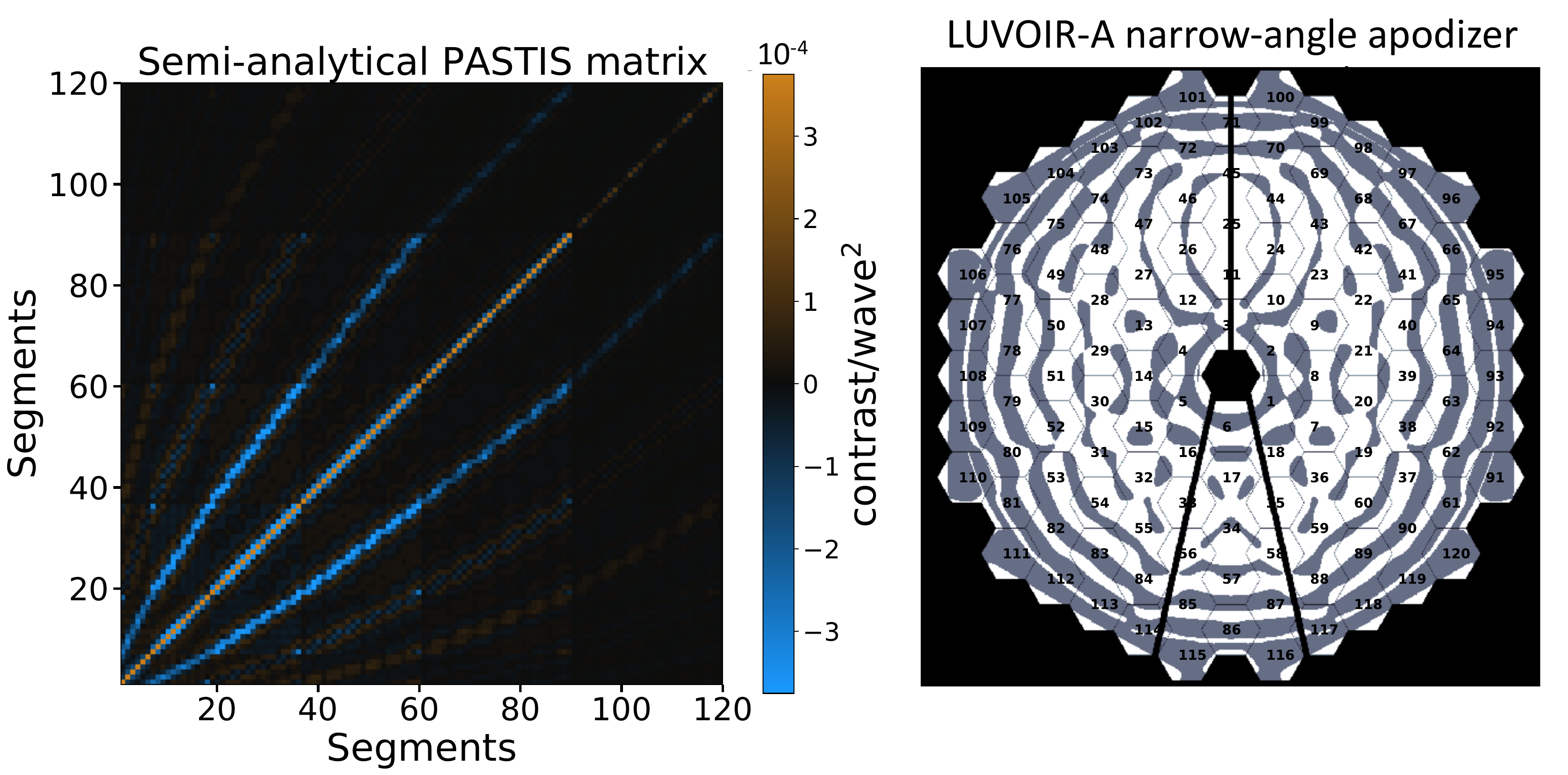}
   \end{tabular}
   \end{center}
   \caption[Numerical matrix] 
   { \label{fig:numerical_matrix} 
    Semi-analytical PASTIS matrix of the 120 segment LUVOIR-A design with the narrow-angle APLC (\textit{left}). The matrix is symmetric by construction and the dark streaks correspond to negative values. The diagonal elements show which segments have more impact on the contrast than others. The outermost segments (65--120) have lower matrix values because of the darker apodization for these segments.  This is clearly visible in the superimposed image of the apodizer on the segments (\textit{right}).}
   \end{figure*}

The semi-analytical PASTIS matrix for the narrow-angle LUVOIR APLC is calculated following Sec.~\ref{subsec:semi-analytic-extension} and shown in Fig.~\ref{fig:numerical_matrix}.

The PASTIS matrix shows how some segments have a higher impact on the final contrast than others. This is visible along the diagonal, which records the contrast contribution from each individual segment alone. For example, segments 65-120 have a lower contrast contribution, as they correspond to the darker areas of the apodizer on the outer two rings of the aperture (see Fig.~\ref{fig:numerical_matrix}, right panel). This effect is also visible on the innermost ring of hexagons. We can also notice streaks of negative values in the matrix in the off-axis areas, as discussed in Sec.~\ref{subsec:semi-analytic-extension}.
   
\begin{figure}
   \begin{center}
   \begin{tabular}{c}
   \includegraphics[width=.67\linewidth]{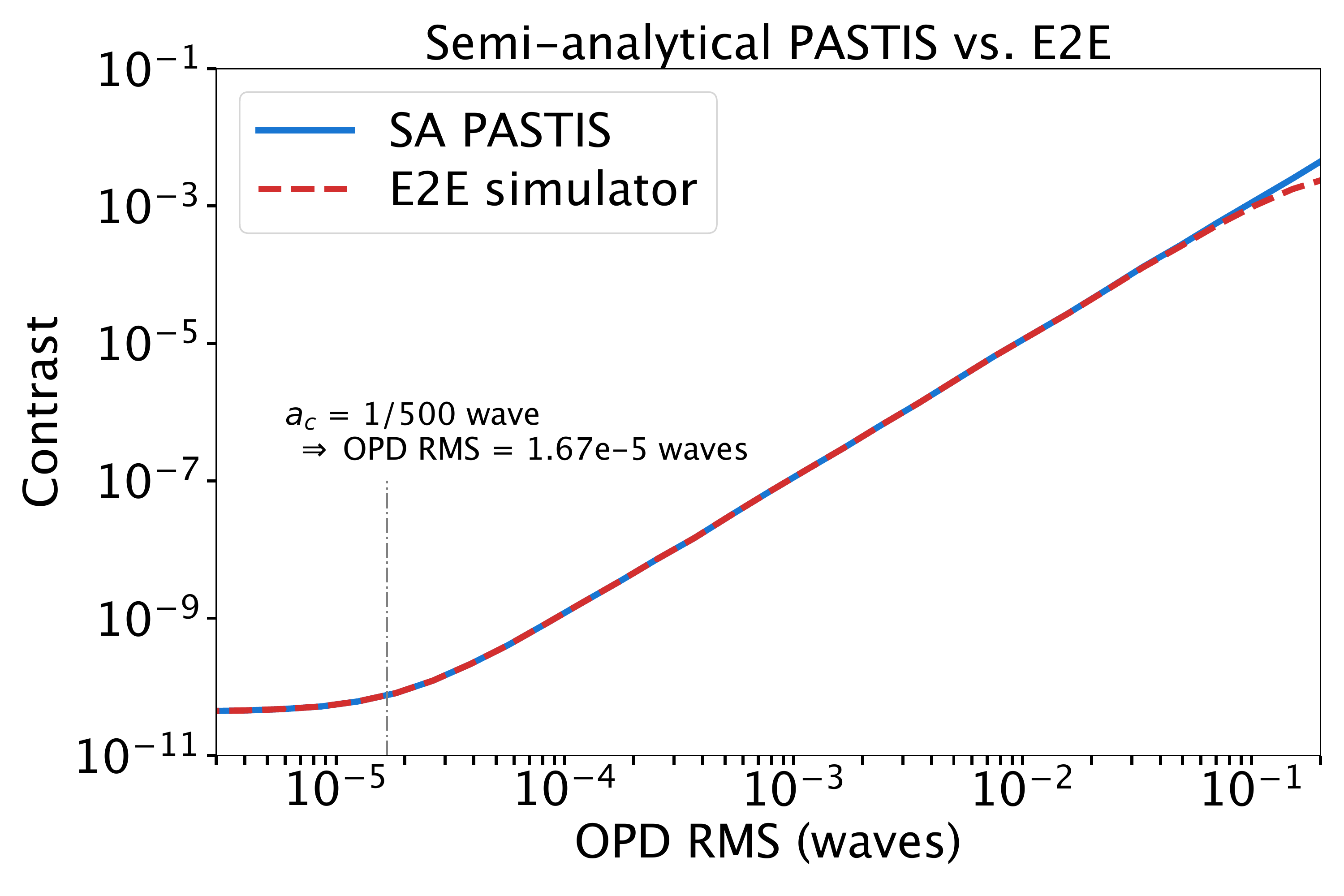}
   \end{tabular}
   \end{center}
   \caption[Contrast validation] 
   { \label{fig:hockeystick} 
    Average dark hole contrast as a function of wavefront error using both an end-to-end simulator (dashed red) and the PASTIS matrix propagation (solid blue). In a hockey stick graph behavior, the contrast is limited by the coronagraph itself at low wavefront errors corresponding to the flattened out curve to the left (at $c_0$). From about $10^{-4}$~waves to $10^{-1}$~waves of WFE RMS, the contrast is limited by segment phasing aberrations. In this range the estimation error of PASTIS is $0.06 \%$ compared to the reference E2E model. A calibration aberration per segment $a_c$ of 1/500 wave, on a 120 segment pupil, translates to a global WFE of $1.67\times10^{-5}$ waves when calibrating the PASTIS matrix diagonal, or $2.36\times10^{-5}$ with two simultaneously aberrated segments, which is in the small aberration regime of the model, just above the coronagraph floor. The curves shown are obtained as the mean of the same 20 random realizations for each RMS value, both for the E2E simulator and the PASTIS propagation. At large wavefront errors (close to 0.1 waves RMS) the linear approximation breaks down and the two curves no longer match perfectly.}
   \end{figure}

We validate the semi-analytical PASTIS matrix by comparing the PASTIS contrast obtained with the matrix formalism of Eq.~\ref{eq:pastis-equation} to the contrast from an E2E simulator using the same inputs. We show the comparison in Fig.~\ref{fig:hockeystick}. The coronagraph floor for this particular APLC design in the absence of aberrations is $4.3 \times 10^{-11}$. The PASTIS model starts to diverge from the E2E calculation at large WFE root-mean-square (RMS) where the linear approximation of the phase breaks down. Note that the choice of $a_c = 1/500$ of the wavelength (used in the presented example, and is on the order of 1\,nm in visible) on a single segment yields a global pupil WFE RMS of $1.67 \times 10^{-5}$ waves, which translates into an average DH contrast just above the coronagraph floor, but keeps it in the small aberration regime.  Here, the accuracy of the semi-analytical matrix approach is significantly higher than that of the fully analytical matrix because the construction of the PASTIS matrix is based on the actual E2E simulation as opposed to a post-calibrated analytical fringe model. 
   
\section{Model inversion and statistical mean contrast derivation}
\label{sec:model-inversion}

Once the PASTIS matrix has been calculated, Eq.~\ref{eq:pastis-equation} gives a fully analytical expression of the dark hole average contrast for any random segment-level aberration $\mathbf{a}$. This makes PASTIS particularly well suited for error budgeting analyses, compared to otherwise computation-intensive Monte-Carlo analyses. 
More interestingly, this analytical model can be inverted to determine the pupil plane aberration vector $\mathbf{a}$ that meets a specific average contrast target $c_t$, using an eigendecomposition of the PASTIS matrix. We also show that the model inversion to obtain the target contrast as a function of eigenmodes is achieved both in a deterministic and statistical sense.

\subsection{Eigendecomposition of the PASTIS model and mode-segment relationship}
\label{subsec:pastis-eigendecomposition}
\begin{figure*}[p]
    \vspace*{-2cm}
    \makebox[\linewidth]{
        \includegraphics[width=1.\linewidth]{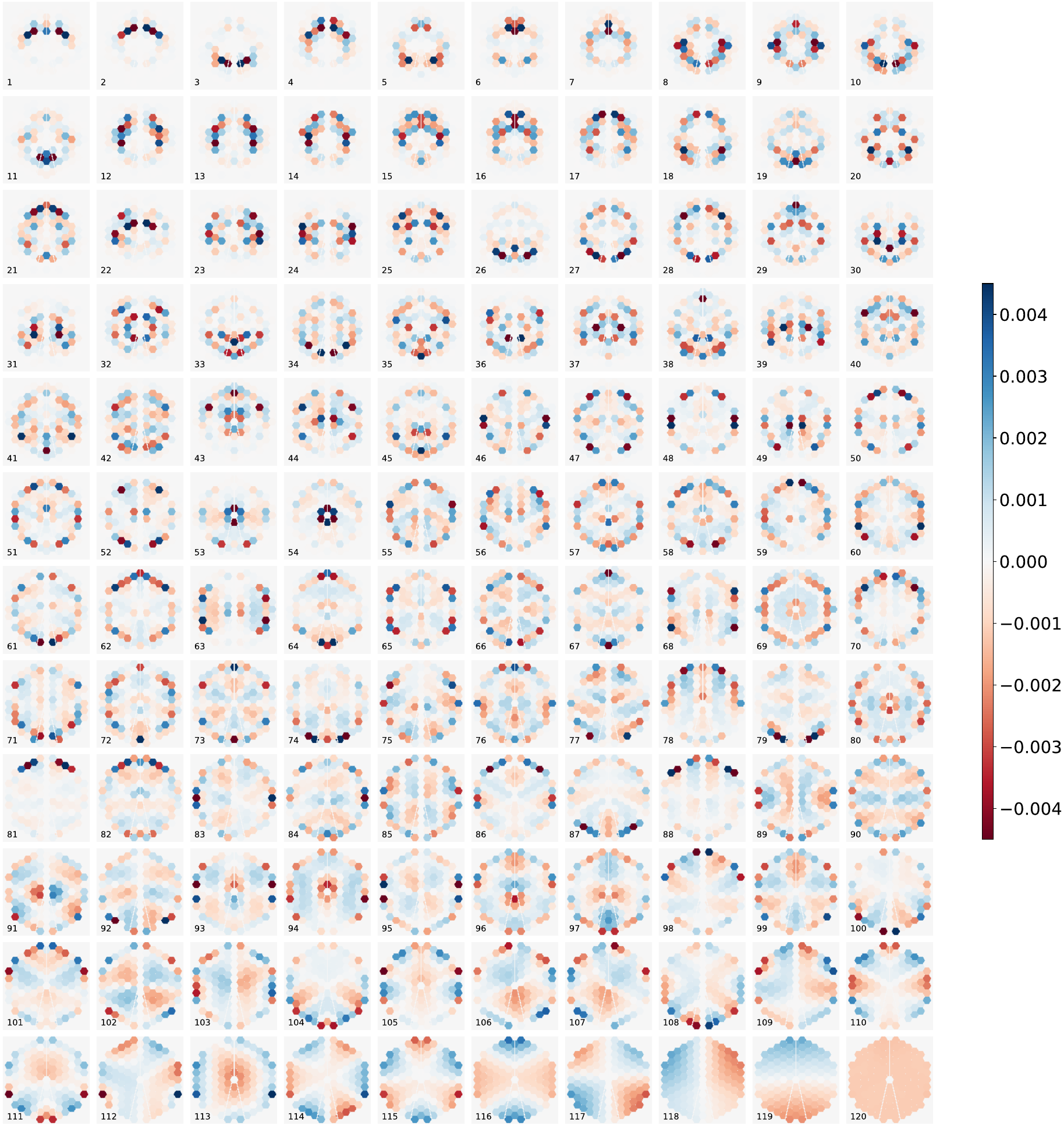}
    }
    \caption{All PASTIS modes for the LUVOIR-A narrow-angle APLC, for local piston aberrations, sorted from highest to lowest eigenvalue. The modes are unitless, showcasing the relative scaling of the segments to each other, and between all modes. They gain physical meaning when multiplied by a mode aberration amplitude $b_p$ in units of wavefront error or phase. Their respective eigenvalues and hence relative impact on final contrast is displayed in Fig.~\ref{fig:eigenvalues}.}
    \label{fig:postage-stamp-modes}
    \end{figure*}

\begin{figure}
   \begin{center}
   \begin{tabular}{c}
   \includegraphics[width=0.67\linewidth]{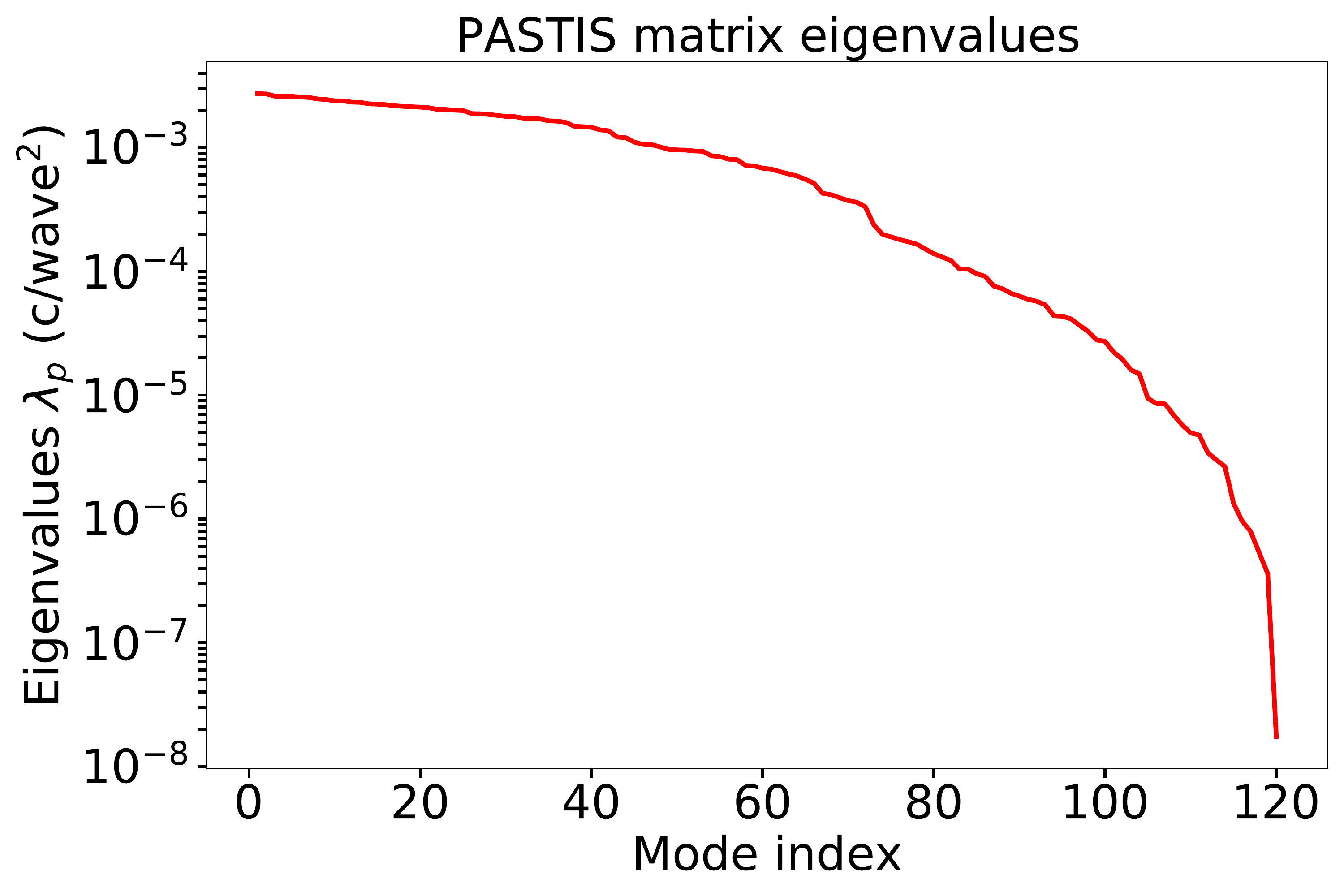}
   \end{tabular}
   \end{center}
   \caption[eigenvalues] 
   { \label{fig:eigenvalues} 
    Eigenvalues, or sensitivity of contrast to mode index $p$, for the piston PASTIS  modes of the LUVOIR-A telescope with the small FPM coronagraph design, shown in Fig.~\ref{fig:postage-stamp-modes}. Note how the PASTIS matrix and modes do not depend on the target contrast, but they do on the choice of telescope geometry and coronagraph, making them the proper modes of the optical system.}
    \end{figure}

\begin{figure}[p]
    \centering
   \includegraphics[width=10cm]{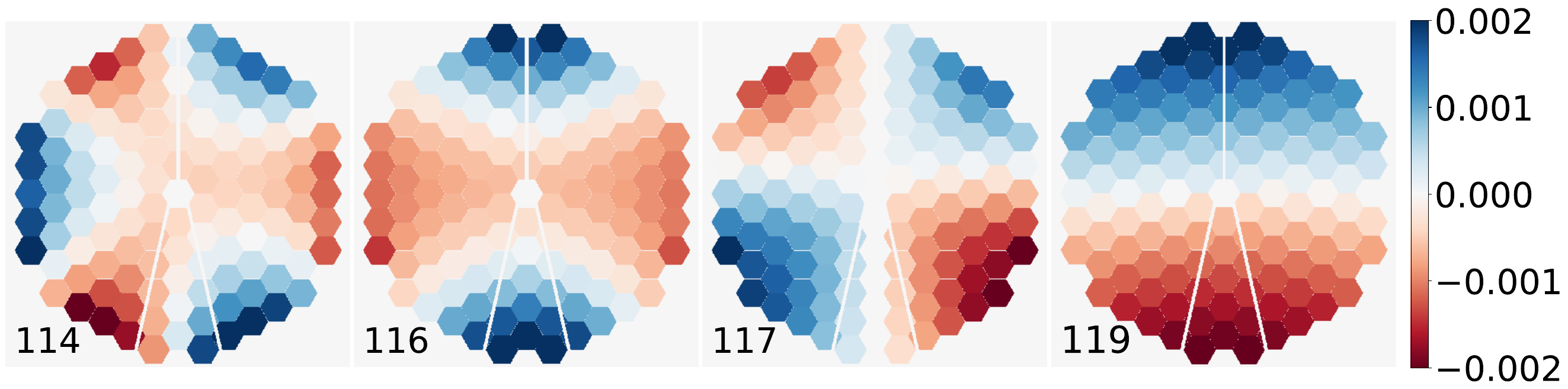}
    \caption[Low Order Modes] 
    {\label{fig:low-order-modes} 
    Low-impact modes with high tolerances for the narrow-angle APLC on the LUVOIR-A telescope, for local piston aberrations. These modes have little impact on the final contrast - they are similar, but not equal, to discretized Zernike modes and the coronagraph rejects them very well by design. }
    \bigskip
   
   \includegraphics[width=10cm]{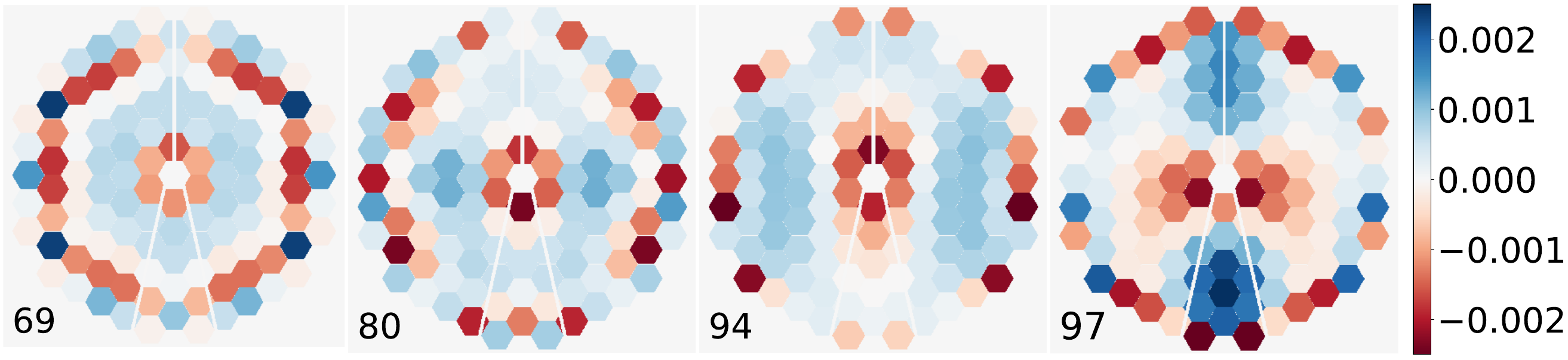}
   \caption[Mid Order Modes] 
   {\label{fig:mid-order-modes} 
    Mid-impact modes with medium tolerances for the narrow-angle APLC on the LUVOIR-A telescope, for local piston aberrations. These modes have medium impact on the final contrast, relatively speaking. These modes show mostly low spatial frequency features except for high spatial frequency components in the parts of the pupil where the apodizer covers most of the segments. }
    \bigskip
   
   \includegraphics[width=10cm]{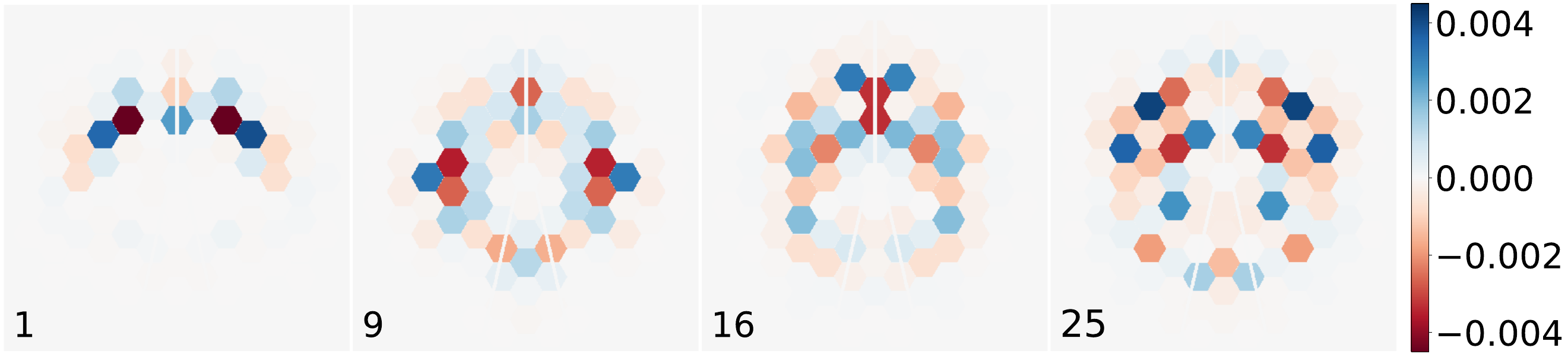}
    \caption[High Order Modes] 
   {\label{fig:high-order-modes} 
    High-impact modes with low tolerances for the narrow-angle APLC on the LUVOIR-A telescope, for local piston aberrations. These modes have the highest impact on the final contrast. They consist entirely of high spatial frequency components in the parts of the pupil where the apodizer (and other pupil plane optics) are the most transmissive.}
    \end{figure}

The PASTIS matrix $M$ is square and symmetric by construction, and therefore diagonalizable. We perform the eigendecomposition:
    \begin{equation}
    M = U  D  U^T,
    \label{eq:eigendecomposition}
    \end{equation}
where U is unitary, hence invertible and $U^{-1} = U^T$. The columns of $U$ are the eigenmodes of the PASTIS matrix $M$, which can be written as column vectors $\mathbf{u}_p$ of $U = \left( \mathbf{u}_1, \mathbf{u}_2, \dots, \mathbf{u}_p, \dots, \mathbf{u}_{n_{modes}} \right)$ and $n_{modes}$ is the total number of eigenmodes (which is equal to the total number of segments $n_{seg}$), indexed by $p$. $D$ is a diagonal matrix whose diagonal elements are the eigenvalues $\lambda_p$ of the matrix $M$; it is the diagonalized PASTIS matrix $D$. The analysis of the eigenmodes $\mathbf{u}_p$ provides information about the critical modes of the system that can be used to place tolerances on segment cophasing and stability. The full set of modes of the LUVOIR-A primary with the narrow-angle APLC is shown in Fig.~\ref{fig:postage-stamp-modes}, and a selection of modes in Figs.~\ref{fig:low-order-modes}, \ref{fig:mid-order-modes}, and \ref{fig:high-order-modes}.
The eigenvalues $\lambda_p$ shown in Fig.~\ref{fig:eigenvalues} indicate how much each mode contributes to the final image contrast if applied to the pupil in their natural normalization, without any imposed weighting. This figure shows that the high-spatial frequency modes (to the left) have a much higher impact than the lower-spatial frequency modes (to the right).

The PASTIS modes $\mathbf{u}_p$ form an orthonormal basis set that allows us to express any arbitrary, segment-based pupil plane aberration $\mathbf{a}$ as a linear combination of the modes $\mathbf{u}_p$ with mode weighting factors $b_p$:
    \begin{equation}
    \mathbf{a} = \sum_{p=1}^{n_{modes}} \mathbf{u}_p b_p.
    \label{eq:lin-combo}
    \end{equation}
This can also be written as:
    \begin{equation}
    \mathbf{a} = U \cdot \mathbf{b},
    \label{eq:basis-transformation}
    \end{equation}
indicating a basis transformation between the mode basis and the segment basis. The inverse basis transformation is thus given by $\mathbf{b} = U^{-1} \mathbf{a}$. This relationship demonstrates the physical equivalence of working in the mode-basis or in the segment basis, as we can transform any expression in one space into an expression of equivalent meaning in the other space. We further explore the physical meaning of the PASTIS modes in Sec.~\ref{subsec:contrast-from-modes} and \ref{subsec:statistical-sigmas}.

\subsection{Contrast as a function of the eigenmodes}
\label{subsec:contrast-from-modes}

\begin{figure}
   \begin{center}
   \begin{tabular}{c}
   \includegraphics[width=11cm]{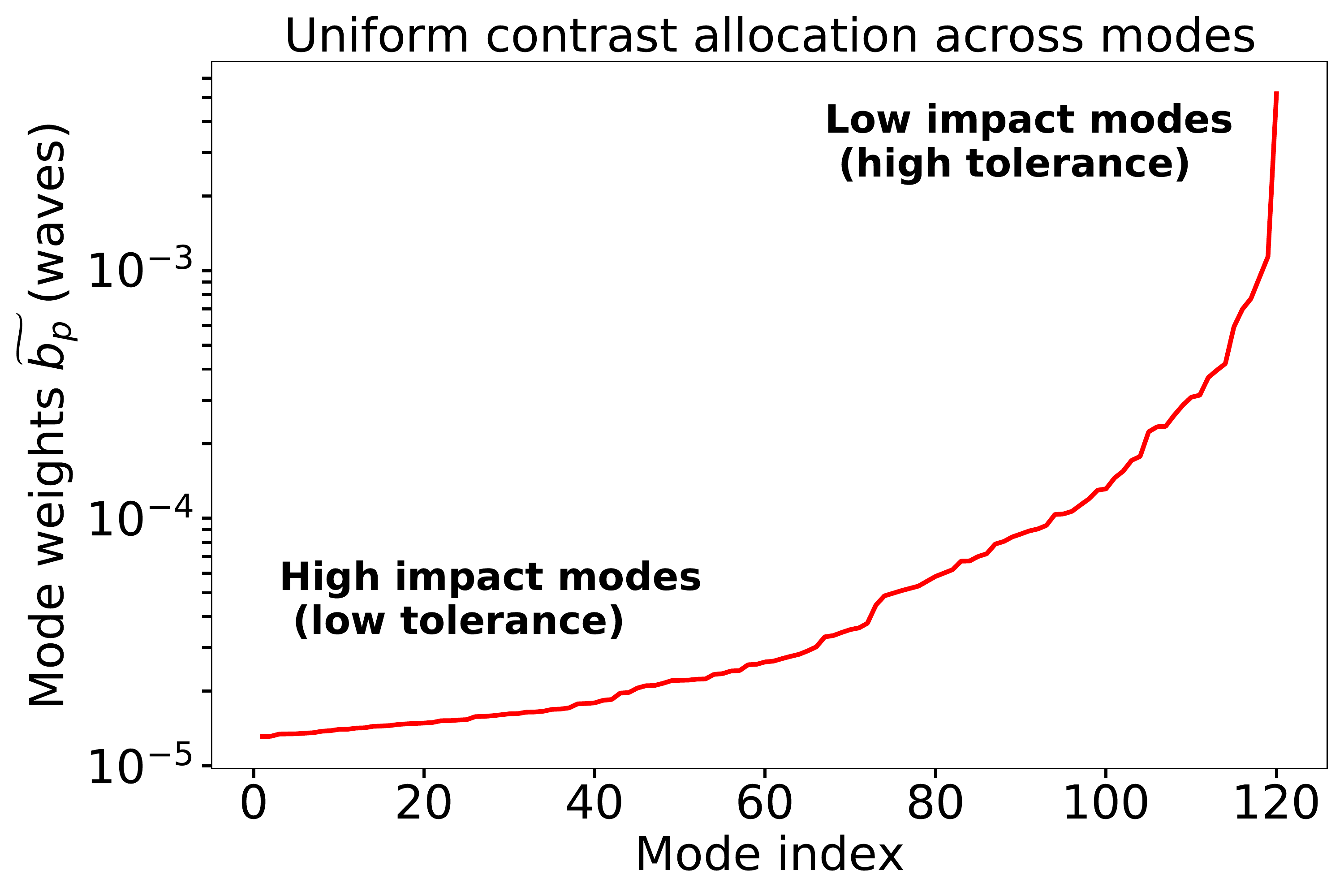}
   \end{tabular}
   \end{center}
   \caption[Sigmas flat error-budget] 
   {\label{fig:sigmas-flat-error-budget} 
    PASTIS mode weights for the uniform contrast allocation across all modes. The low-index modes to the left, which correspond to high spatial frequencies, have a lower WFE tolerance than the low spatial frequency modes with high index to the right. These mode amplitudes are inversely proportional to the eigenvalues associated with each mode (Eq.~\ref{eq:calc-sigma-uniform}), and they scale the modes such that each of them contributes the same contrast $c_p$ to the overall target contrast. The cumulative contrast response of the modes multiplied by these weights is shown in Fig.~\ref{fig:cumulative-contrast}.}
   \end{figure}

 The mode weights $\mathbf{b}$ will depend on how much each individual mode contributes to the final contrast, and their associated eigenvalue. Inserting Eq.~\ref{eq:basis-transformation} into Eq.~\ref{eq:pastis-equation} allows us to define this relationship:
    \begin{equation}
    \begin{aligned}
    c - c_0 &= (U \mathbf{b})^T M (U \mathbf{b})\\
    &= \mathbf{b}^T U^T M U \mathbf{b}\\
    &= \mathbf{b}^T D \mathbf{b},\\
    \label{eq:pastis-equation-mode-basis}
    \end{aligned}
    \end{equation}
and finally:
    \begin{equation}
    c - c_0 = \sum_{p}^{n_{modes}} b_p^2 \lambda_p.
    \label{eq:get-bs-from-matrices}
    \end{equation}
The final contrast is therefore the sum of all squared mode weights, multiplied by their respective eigenvalue. Since the modes contribute independently to the final contrast (they are orthonormal by construction), we can define a per-mode contrast as:
    \begin{equation}
    c_p = b_p^2 \lambda_p,
    \label{eq:simple-cp}
    \end{equation}
and obtain that the total contrast is the sum of all individual contrast contributions:
     \begin{equation}
     c = c_0 + \sum_{p=1}^{n_{modes}} c_p.
     \label{eq:target-contrast}
    \end{equation}
We can then find the p-th mode weight that gives the allocated contrast contribution $c_p$ as:
    \begin{equation}
    b_p = \sqrt{\frac{c_p}{\lambda_p}}.
    \label{eq:calc-sigma}
    \end{equation}
Eq.~\ref{eq:calc-sigma} gives the weighting factor for each PASTIS mode when it has a particular contrast contribution $c_p$. We can illustrate this expression by calculating the mode weights corresponding specifically to a uniform contrast contribution of the overall target contrast $c_t$ over all modes, $c_p = (c_t - c_0)/n_{modes}$. Then we calculate them as
    \begin{equation}
    \widetilde{b_p} = \sqrt{\frac{c_t - c_0}{n_{modes} \cdot \lambda_p}},
    \label{eq:calc-sigma-uniform}
    \end{equation}
where the $\widetilde{b_p}$ is the particular set of mode weights in the case of a uniform contrast allocation across all modes. The resulting mode weights $\widetilde{\mathbf{b}}$ for a total contrast allocation of $c_t = 10^{-10}$ are shown in Fig.~\ref{fig:sigmas-flat-error-budget}.

\begin{figure}
   \begin{center}
   \begin{tabular}{c}
   \includegraphics[width=11cm]{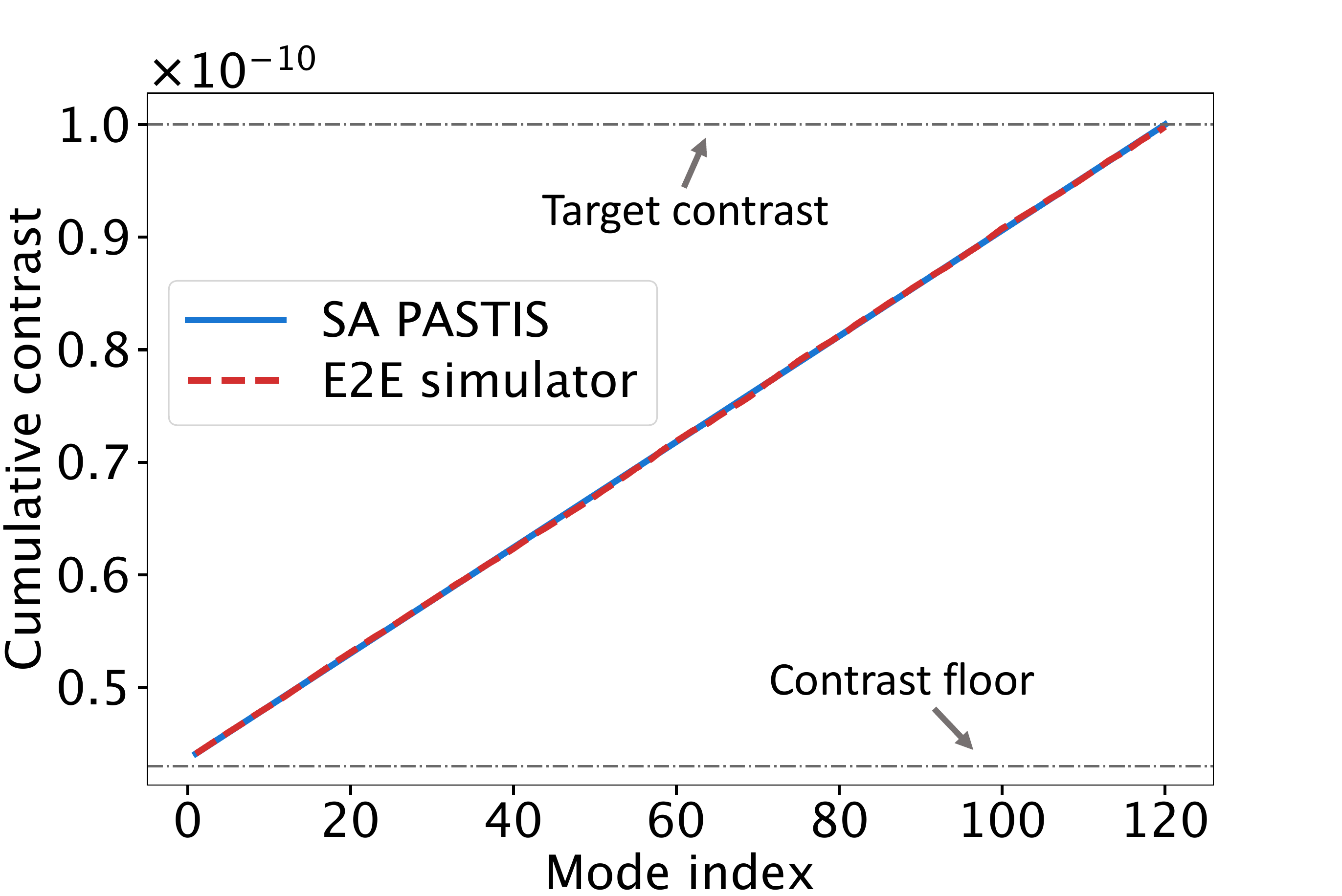}
   \end{tabular}
   \end{center}
   \caption[CumulativeContrast] 
   {\label{fig:cumulative-contrast} 
    Cumulative contrast from all PASTIS modes when allocating the total contrast uniformly across all modes (see Sec.~\ref{subsec:contrast-from-modes}). For instance, the measured contrast corresponding to the first 60 accumulated weighted modes is about $0.7\times10^{-10}$. We multiply all modes by their respective mode amplitude $\widetilde{b_p}$ and propagate them cumulatively to the image plane, both with the E2E simulator (dashed red) and the PASTIS propagation (solid blue). Without any of the modes applied, we get the contrast floor from the coronagraph $c_0$, while application of all modes together yields the requested target contrast, here $c_t = 10^{-10}$. Each mode is allocated an equal contrast contribution $c_p$ to the final contrast, which results in a linear cumulative contrast curve. Note how neither line starts at the coronagraph floor because the lowest-index mode already adds a contrast contribution on top of the baseline contrast. The corresponding PASTIS mode weights to obtain this uniform allocation of contrast per mode, is shown in Fig.\ref{fig:sigmas-flat-error-budget}.}
   \end{figure}

In Fig.~\ref{fig:cumulative-contrast} we confirm the validity of the mode weights $\widetilde{b_p}$ by showing the average dark hole contrast from an end-to-end propagation of the cumulative wavefront error for all modes. The linearity of the plot, as well as the end value at the target contrast $c_t$ validates the uniform contrast allocation to each PASTIS mode from Eq.~\ref{eq:calc-sigma-uniform}. 

\subsection{Statistical mean of the contrast from mode amplitudes}
\label{subsec:statistical-sigmas}

\begin{figure}
   \begin{center}
   \begin{tabular}{c}
   \includegraphics[width=10cm]{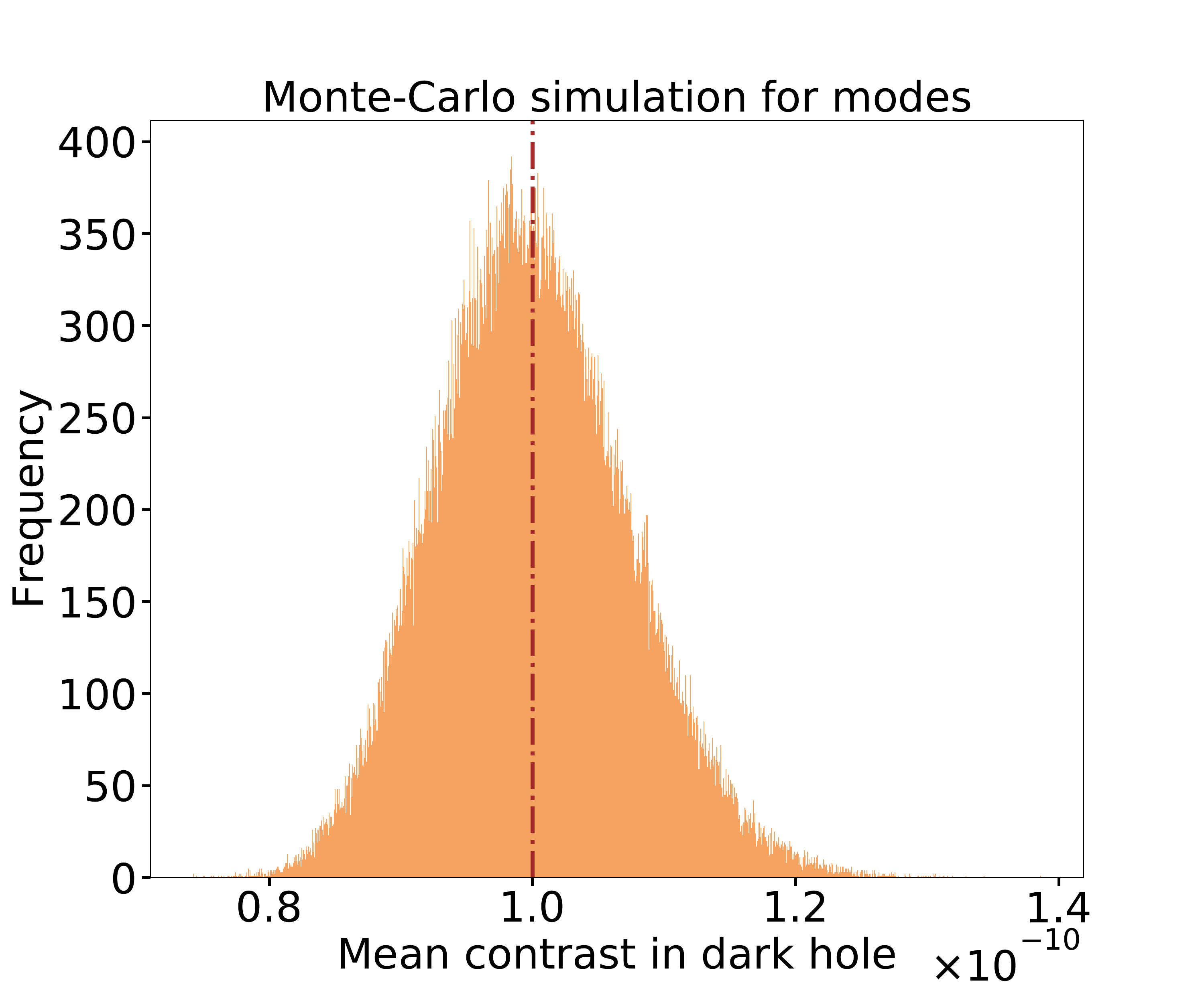}
   \end{tabular}
   \end{center}
   \caption[MC-for-sigmas] 
   {\label{fig:MC-for-sigmas} 
    alidation of the uniform contrast allocation across PASTIS modes with an E2E Monte-Carlo simulation, drawing random sets of mode weights from a normal distribution with standard deviations $\bm{\sigma}$ (Eq.~\ref{eq:sigma-as-standard-deviation}),  corresponding to an equal contrast allocation per mode (Eq.~\ref{eq:calc-sigma-uniform}) with weights $\mathbf{\widetilde{b}}$. The overall WFE of one realization is the sum of all weighted modes for each set, and is propagated in an E2E simulation. The histogram represents 100,000 realizations of the average dark hole contrast for a target contrast of $10^{-10}$.}
    \end{figure}

In this section, we analyze the properties of the model in the statistical sense to prepare a framework for the segment-level error budget in Sec.~\ref{sec:segment-error-budget}. We extend the formalism from purely deterministic mode weights $\mathbf{b}$ to random variables. We obtain the statistical mean contrast by substituting Eq.~\ref{eq:simple-cp} into Eq.~\ref{eq:target-contrast} and taking the mean:
    \begin{equation}
    \langle c\rangle - c_0 = \sum_p^{n_{modes}} \langle b_p^2\rangle \lambda_p.
    \label{eq:statistical-sigmas}
    \end{equation}
Assuming zero-mean normal distributions of the PASTIS modes, we can readily identify their variance as
    \begin{equation}
    \sigma_p^2 = \langle b_p^2\rangle.
    \label{eq:sigma-as-standard-deviation}
    \end{equation}
We verify this for the uniform contrast allocation per mode (Eq.~\ref{eq:calc-sigma-uniform}) in an E2E Monte-Carlo simulation where we draw random samples of the PASTIS mode coefficients $\mathbf{\widetilde{b}}$, following zero-mean normal distributions $\mathcal{N}$ with standard deviations $\sigma_p$, according to the uniform contrast allocation from Eq.~\ref{eq:calc-sigma-uniform}:
    \begin{equation}
    \mathbf{b} = (\mathcal{N}(0,\sigma_1),\ \mathcal{N}(0,\sigma_2),\ \ldots,\ \mathcal{N}(0,\sigma_p)).
    \label{eq:b-normal-distribution}
    \end{equation}
Fig.~\ref{fig:MC-for-sigmas} shows the result of such a Monte-Carlo simulation where we sum each randomly weighted, individual set of modes to a unique wavefront map, propagate it to the image plane with the E2E simulator and measure the spatial average contrast in the dark hole. We validate that the statistical mean of all these contrast values is the target contrast $c_t$ for which we calculated the vector of mode standard deviations $\bm{\sigma}$ in the first place, in this case $10^{-10}$.

In summary, PASTIS provides an analytical model to go from a set of segment aberrations to the DH average contrast in the coronagraphic image. The calculation of the PASTIS matrix eigenmodes allows to invert this model: we can set a target contrast and allocate WFE amplitudes to each of the system's eigenmodes to reach that target contrast. Since they form an orthonormal set of modes, their individual contrast contributions add independently, each with its own sensitivity, as a fraction of the average contrast in Sec.~\ref{subsec:contrast-from-modes}, on top of their statistical description in Sec.~\ref{subsec:statistical-sigmas}. These optical modes contain the full information of the image formation system, including the apodizer, coronagraph, primary geometry and other optical components in the system, and are distinct from and unaware of the mechanical behavior of the telescope. By this nature, we can use them to understand the fundamental limitations for high contrast imaging with a segmented aperture, which will in sequence be further constrained by thermo-mechanical properties of the telescope as we describe in Sec.~\ref{sec:segment-error-budget}.

\section{Segment-level tolerance statistics}
\label{sec:segment-error-budget}

Sec.~\ref{sec:model-inversion} details how we can construct mode-based requirements that satisfy a target on the spatial average contrast in the image plane. This relationship is not only valid in a deterministic sense (relating a mode amplitude directly to its contrast contribution), but also in the statistical sense (relating the standard deviation of random modes to the overall dark hole contrast). The PASTIS modes form an orthonormal basis where each mode contributes independently to the dark hole average contrast: the contrast of a sum of weighted modes is equal to the sum of the contrast contributions for all weighted modes. The PASTIS matrix $M$ and its eigenmodes fully describe the optical propagation through the system in terms of WFE effects from a segmented aperture on the average contrast in the DH with a coronagraph. However, they do not contain any information about thermal or mechanical effects necessary to describe the final performance of a given segmented observatory. A sound framework to develop error budgets on segmented apertures therefore requires the combination of both the optical response of the telescope and coronagraph (encapsulated in the PASTIS modes), and the thermo-mechanical response of the telescope and the observatory (encapsulated in the segment aberration covariance matrix).

The goal of this section is to combine the information about the imaging formation through the coronagraph on a segmented mirror with the thermo-mechanical properties of the observatory, in order to establish requirements to reach a target DH contrast.

\subsection{Statistical mean contrast and its variance in segmented coronagraphy}
\label{subsec:mean-and-variance-segmented-coronagraphy}

We can calculate the statistical mean contrast of the DH spatial average directly from Eq.~\ref{eq:pastis-equation}, exploiting the fact that the trace of a scalar is the scalar itself, and that $\tr(AB) = \tr(BA)$: 
   \begin{equation}
    \begin{aligned}
    \langle c \rangle &= c_0 + \langle \mathbf{a}^T M \mathbf{a} \rangle = c_0 + \langle \tr(\mathbf{a}^T M \mathbf{a}) \rangle \\
    &= c_0 + \langle \tr(M \mathbf{a} \mathbf{a}^T) \rangle = c_0 + \tr(M \langle \mathbf{a} \mathbf{a}^T \rangle),
    \end{aligned}
    \label{eq:derive-contrast-with-trace}
    \end{equation}
and finally:
    \begin{equation}
    \langle c \rangle = c_0 + \tr(M C_a),
    \label{eq:avg-contrast-from-trace}
    \end{equation}
where $C_a$ is the $n_{seg} \times n_{seg}$ segment covariance matrix, containing the as-built thermo-mechanical correlations between segments. 
Eq.~\ref{eq:avg-contrast-from-trace} allows us to calculate the statistical mean of the average DH contrast directly from the knowledge of the segment covariance matrix, no matter if there is correlation between the segments or not, combining the imaging properties of the high contrast imaging system, contained in $M$, with the thermo-mechanical behavior of the instrument contained in $C_a$.

Similarly, we can derive an analytical expression for the variance $\Var(c)$ of the DH contrast. Assuming that $\mathbf{a}$ follows a zero-mean Gaussian distribution, the variance for Eq.~\ref{eq:pastis-equation} takes the very simple form\cite[Theorem~5.2c]{Rencher2008}:
    \begin{equation}
    \Var (c) = 2 \tr [(M C_a)^2].
    \label{eq:var-of-c}
    \end{equation}
    
These two equations provide an unambiguous closed form derivation of the mean contrast and its variance from the optical model of the imaging system (encapsulated in $M$), and from the thermo-mechanical properties of the telescope (captured by $C_a$). The PASTIS matrix $M$ knows nothing of the thermo-mechanical effects of the observatory and is obtained by diffractive modelling of the coronagraph, while the segment covariance matrix comes from thermal and mechanical modelling of the observatory and is completely detached from the image formation system of the telescope. The two matrices together ($M$ and $C_a$) fully describe the statistical response of the coronagraph system to a particular WFE allocation on segments and therefore allow to establish a set of top-level requirements on segment tolerances for an observatory.

The enabling aspect of Eqs.~\ref{eq:avg-contrast-from-trace} and \ref{eq:var-of-c} for segment-level tolerancing is that the trace is invariant under a basis transformation. It follows that if either one of the two matrices, the PASTIS matrix or the thermo-mechanical covariance matrix, is expressed in its diagonal basis, the expressions for the contrast mean and variance simplify greatly, as we show in the following sections. The segment tolerancing can thus be achieved either by diagonalizing M, or by doing so with $C_a$. We have treated the case of diagonalizing the PASTIS matrix $M$ in Sec.~\ref{sec:model-inversion}, where we describe the analytical framework for segmented telescope tolerancing in the diagonal basis that most naturally describes the optical sensitivity of the system to the DH contrast. In the following two sections, we turn to a basis that diagonalizes the segment covariance matrix instead, permitting us to perform the tolerancing on appropriate system modes.

\subsection{Uncorrelated segment-level requirements}
\label{subsec:segment-level-requirements}

In finding a diagonal basis for the thermo-mechanical matrix, the easiest case is when $C_a$ is already diagonal, which physically corresponds to independent segments on the primary mirror. In this case, the diagonal elements of $C_a$, namely the segment variances $\langle a_k^2 \rangle$, fully describe the effect of the primary mirror segments on the DH contrast, and the statistical mean of the contrast in Eq.~\ref{eq:avg-contrast-from-trace}, $\langle c\rangle$, finds a simple expression similar to Eq.~\ref{eq:statistical-sigmas}:
    \begin{equation}
    \langle c \rangle = c_0 +  \sum_k^{n_{seg}} m_{kk} \langle a_k^2 \rangle.
    \label{eq:mean-contrast-for-independent-segments}
    \end{equation}
Similarly to the mode-based error budget presented in Sec.~\ref{sec:model-inversion}, we now want to find a segment-based error budget to formulate the WFE limits on each segment that reach a specific statistical mean target contrast $c_t = \langle c \rangle$. Turning to a statistical mean contrast allows us to define a similar allocation of contrast contributions to all segments as we did statistically (and deterministically) in the PASTIS mode basis (Eq.~\ref{eq:target-contrast}). The most straightforward way of doing this is to allocate the target contrast equally to all segments:
    \begin{equation}
    \langle a_k^2\rangle m_{kk} = \frac{\langle c \rangle - c_0}{n_{seg}}.
    \label{eq:segment-requirement}
    \end{equation}
If we define $\mu_k$ as the standard deviation of the WFE on the $k$-th segment, comparably to Eq.~\ref{eq:sigma-as-standard-deviation}:
    \begin{equation}
    \mu_k^2 = \langle a_k^2\rangle,
    \label{eq:mu-as-standard-deviation}
    \end{equation}
then by combining the three previous equations we obtain the per-segment WFE requirement  for this particular contrast allocation:
    \begin{equation}
    \mu_k^2  = \frac{\langle c \rangle -  c_0}{ n_{seg} m_{kk}}.
    \label{eq:single-mus}
    \end{equation}
    
The expression in Eq.~\ref{eq:single-mus} lets us calculate a per-segment requirement for all individual segments in the pupil of a coronagraphic instrument, given a statistical mean target contrast. The main assumption for this is that the segments are independent from each other, and that we have access to the statistical mean value of the contrast. While the mean contrast is easily measurable on images through averaging, this might not be the case for an ultra-stable facility like LUVOIR. However, the statistical mean contrast is an important quantity to perform segment-level WFE tolerancing, especially with regards to mirror manufacturing. 
   
In a more physical sense, we know that the intensity or contrast is proportional to the variance of the WFE. Therefore, the total final contrast over the full pupil is proportional to the sum of the segment variances, which is also proportional to the sum of the contrast for all segments, conforming with Parseval's theorem. We validate this independent segment-level error budget for three different LUVOIR coronagraphs in Sec.~\ref{sec:APPLICATION-TO-LUVOIR}.

\subsection{Case of correlated segments}
\label{subsec:non-independent-segments}

While the assumption of statistically independent segments brings insights into WFE tolerances of segmented mirrors for coronagraphic imaging, it is not general enough to encompass all possible modes for such telescopes where large-scale thermo-mechanical drifts occur (for example, backplane ``flapping" mode around the folding motion of the primary mirror).
Here we discuss extensions of the PASTIS approach to the case of correlated segments. 

In the case of independent segments, the covariance matrix $C_a$ of the aberration vector $\mathbf{a}$ is a simple diagonal matrix holding the segment variances $\mu_k^2$.
However, the covariance matrix $C_a$ is no longer diagonal for correlated segments, because of mechanical coupling for example due to large-scale backplane deformations. 
In this case, the statistical mean contrast and its variance remain analytically computable with Eqs.~\ref{eq:avg-contrast-from-trace} and \ref{eq:var-of-c}. The tolerancing can be done by performing an eigendecomposition on $C_a$, which will diagonalize it and provide an orthonormal set of of eigenmodes that describe the mechanical perturbations of the telescope system, which is also known as the Karhunen-Lo\`eve basis. By writing $C_a = V C_{th} V^T$, we obtain the PASTIS matrix in this new basis, $M'$, through the transformation matrix $V$, where $M' = V M V^T$ ($m'_{kk} \in M'$) still describes the optical properties of the system. In this basis, Eq.~\ref{eq:avg-contrast-from-trace} takes the form :
    \begin{equation}
    \langle c \rangle = c_0 + \tr(M' C_{th}).
    \label{eq:avg-contrast-diagonal}
    \end{equation}
Since the thermo-mechanical covariance matrix $C_{th}$ is diagonal, we identify its diagonal elements as the thermo-mechanical mode variances $s^2_k$. Like for Eqs.~\ref{eq:statistical-sigmas} and \ref{eq:mean-contrast-for-independent-segments}, this simplifies Eq.~\ref{eq:avg-contrast-diagonal} yet again to:
    \begin{equation}
    \langle c \rangle = c_0 +  \sum_k^{n_{th}} m'_{kk}  s^2_k,
    \label{eq:mean-contrast-for-independent-segments-diagonal}
    \end{equation}
which allows us to make a reasonable allocation of  contrast contributions across all $n_{th}$ thermo-mechanical eigenmodes. In the same way like Eq.~\ref{eq:single-mus} calculates a per-segment variance in the basis of independent segments, we can use Eq.~\ref{eq:mean-contrast-for-independent-segments-diagonal} to tolerance the per-mode variances $s^2_k$ to any given target contrast, albeit this time for individual mechanical eigenmodes. Similarly, a transformed expression can be found for the contrast variance in Eq.~\ref{eq:var-of-c}:
    \begin{equation}
    \Var (c) = 2 \tr [(M' C_{th})^2].
    \label{eq:var-of-c-diagonal}
    \end{equation}
In this most general case of correlated segments, the knowledge of the segment-level covariance matrix $C_a$ and its diagonal eigenbasis $C_{th}$ supersedes the simpler description in terms of segment-level variances that is only possible in the uncorrelated case (Sec.~\ref{subsec:segment-level-requirements}). It allows us to  express the mean contrast explicitly as a function of variances of thermo-mechanical modes that can be toleranced in a similar fashion to what was done in Sec.~\ref{sec:model-inversion}.

We have presented a quantitative, fully analytical method to calculate segment-level tolerances for a high-contrast instrument on a segmented aperture telescope. These follow directly from the PASTIS matrix for which we provided a new, semi-analytical way for its calculation that exploits a numerical simulator to compute the effects of the segments on the intensity in the image plane. We encode the optical and thermo-mechanical properties of the observatory separately, with the PASTIS matrix $M$ and the segment covariance matrix $C_a$, which when put together allow for the analytical calculation of the expected mean contrast and its variance. These equations are invariant under a basis transformation, which permits us to find an appropriate diagonalized basis in order to derive individual WFE tolerances. This can either be done by diagonalizing $M$, as we showed in Sec.~\ref{sec:model-inversion}, or by finding a diagonal basis for $C_a$. A special case is given if $C_a$ is naturally diagonal due to independent segments on the segmented mirror; in this case, we derive a per-segment requirement map by following the analytical framework set forth in Sec.~\ref{sec:model-inversion}. In the more general case of correlations between the segments, due to thermo-mechanical properties of the telescope, we diagonalize $C_a$ and use the same analysis principles in the Karhunen-Lo\`eve basis of $C_a$, which allows us to calculate per-mode WFE tolerances. While the deformation matrix $C_a$ will be acquired through thermo-mechanical modelling and can include thermal, vibrational or gravitational perturbations, the PASTIS matrix $M$ and the PASTIS modes give insight into the purely optical properties of the observatory, and the sensitivity of the optical system to contrast.

In the next section, we validate the segment-level error budget in the case of independent segments (Sec.~\ref{subsec:segment-level-requirements}) for three different APLC designs for LUVOIR.

\section{Application to LUVOIR WFE tolerancing}
\label{sec:APPLICATION-TO-LUVOIR}

The LUVOIR study\cite{TheLUVOIRTeam2019} has two point-design cases (LUVOIR-A and LUVOIR-B), respectively 15\,m and 8\,m in diameters, each designed as a TMA and containing a suite of scientific instruments that include coronagraphs. The LUVOIR-A coronagraphic instrument\cite{Pueyo2017} includes a suite of three numerically optimized APLC coronagraphs\cite{Por2020progressive} with focal plane mask diameters that maximize the exo-Earth yield in both detection and characterization\cite{Stark2015, Stark2019}. The optical train of an APLC\cite[Fig.~5]{Leboulleux2018jatis} contains an apodizer in the pupil plane that modulates the optical beam in amplitude, a focal plane mask occulting the on-axis point-spread-function (PSF) core, and a Lyot stop in the subsequent pupil plane that blocks the light diffracted at the focal plane mask (FPM)\cite{Soummer2003, N'Diaye2015, N'Diaye2016}. Of the three LUVOIR APLCs, we used the smallest FPM coronagraph, or narrow-angle coronagraph, for theory validation in the previous sections. It is typically used for spectroscopic characterization in the wavelength band where molecular oxygen and water can be detected ($0.76\, \mu m$ and $0.94\, \mu m$). Planet detection can however be performed at shorter wavelengths (e.g. around $0.5\, \mu m$) where a given angular size corresponds to a larger inner working angle in diffraction resolution units ($\lambda/D$). This larger inner working angle corresponds to a larger FPM, and a larger FPM allows for apodizer designs with a higher throughput and a more robust coronagraph design, which is where the trade-off between the three designs (narrow-, medium- and wide-angle) is made. The three LUVOIR-A APLC designs are shown in the top row of Fig.~\ref{fig:segment-maps}. The corresponding FPM have radii of 3.50, 6.82 and 13.38 $\lambda/D$ respectively, followed by a hard edge annular Lyot stop, whose inner and outer diameters are 12.0\% and 98.2\% of the circumscribed diameter of the apodizers. The resulting coronagraphic dark hole sizes are 3.4--12, 6.7--23.7 and 13.3--46.9 $\lambda/D$ respectively, with a coronagraph floor $c_0$ of $4.3 \times 10^{-11}$, $3.9 \times 10^{-11}$ and $3.9 \times 10^{-11}$ for the three designs.

In this section, we present a full analysis to obtain segment requirements for these three LUVOIR APLCs, and validate the results by performing Monte-Carlo simulations with an E2E simulator. We also take a deeper look into the narrow-angle APLC by analyzing the PASTIS mode-based decomposition of the individual-segment requirements. This monochromatic analysis was performed at a wavelength of 500~nm, which is the lower limit wavelength for the LUVOIR coronagraphs and where we expect to detect planets.

\subsection{Segment requirements and Monte-Carlo simulations for three APLC designs}
\label{subsec:MC-on-LUVOIR-segments}

\begin{figure*}
   \begin{center}
   \begin{tabular}{@{}c@{}}
   \includegraphics[width=\linewidth]{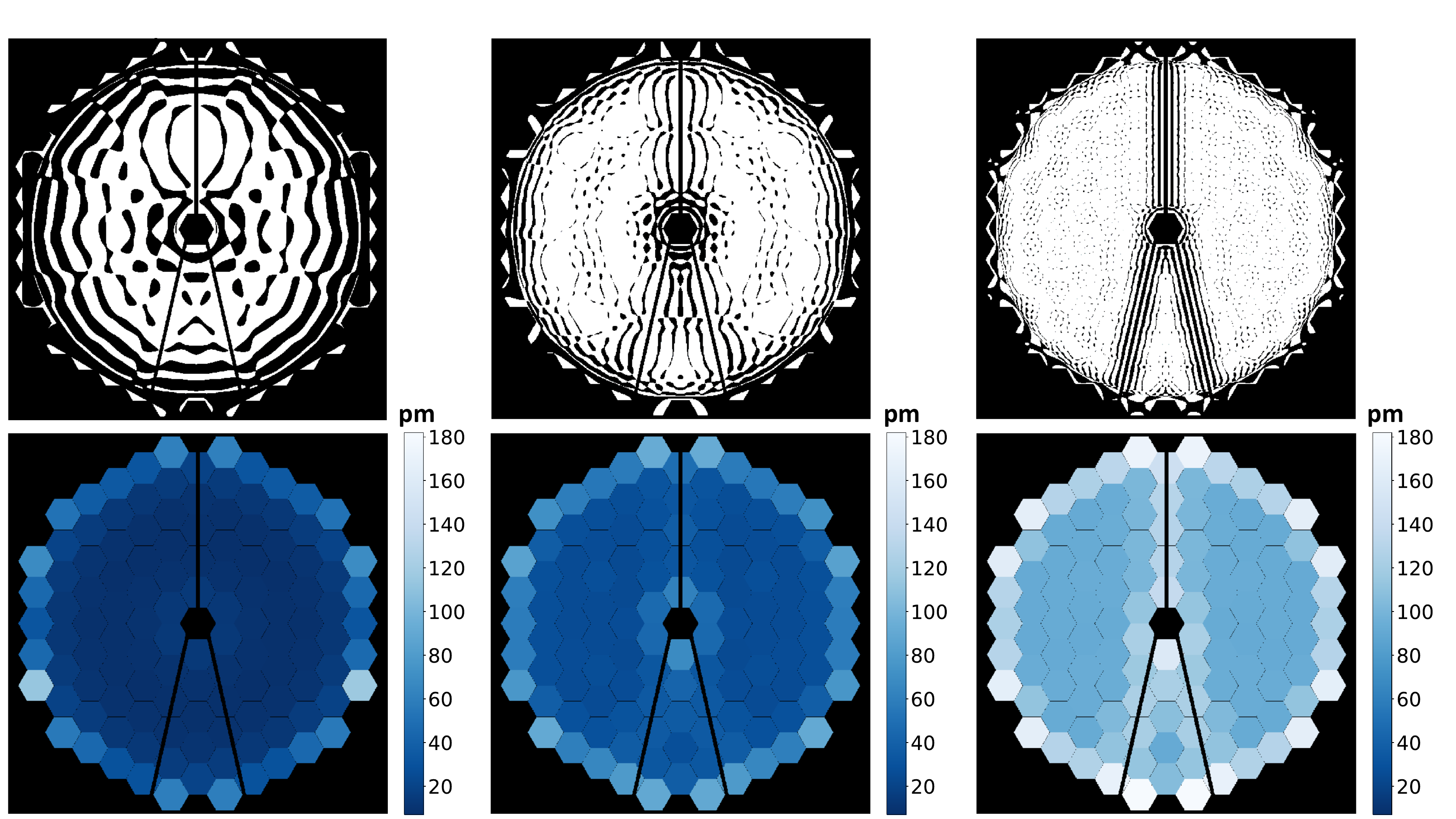}
   \end{tabular}
   \end{center}
   \caption[segment-map-small] 
   {\label{fig:segment-maps} 
    \textit{Top:} The three baseline apodizer designs for LUVOIR-A, a narrow-angle (left), medium-angle (middle) and wide-angle (right) mask (details see Sec.~\ref{sec:APPLICATION-TO-LUVOIR}). \textit{Bottom:} Segment tolerance maps for narrow-angle (left), medium-angle (middle) and wide-angle (right) APLC designs on LUVOIR-A for a target contrast of $c_t = 10^{-10}$, at a wavelength of 500~nm. All three tolerance maps are shown on the same scale. Note how each segment value denotes the standard deviation of a zero-mean normal distribution from which the segment aberrations in WFE RMS are drawn. The minimum and maximum values of these maps are, from left to right: 7 and 116 pm, 25 and 93 pm, and 92 and 181 pm.}
   \end{figure*}

We first calculate the PASTIS matrix for each of these three APLC designs, according to the methodology described in Sec.~\ref{subsec:semi-analytic-extension}. We can then establish a segment-level error budget in the assumption of uncorrelated segments, according to Eq.~\ref{eq:single-mus}. In Fig.~\ref{fig:segment-maps}, we show the resulting segment requirement maps for a target contrast of $c_t = 10^{-10}$ and for all three APLC designs. 

\begin{figure}
   \begin{center}
   \begin{tabular}{c}
   \includegraphics[width=9cm]{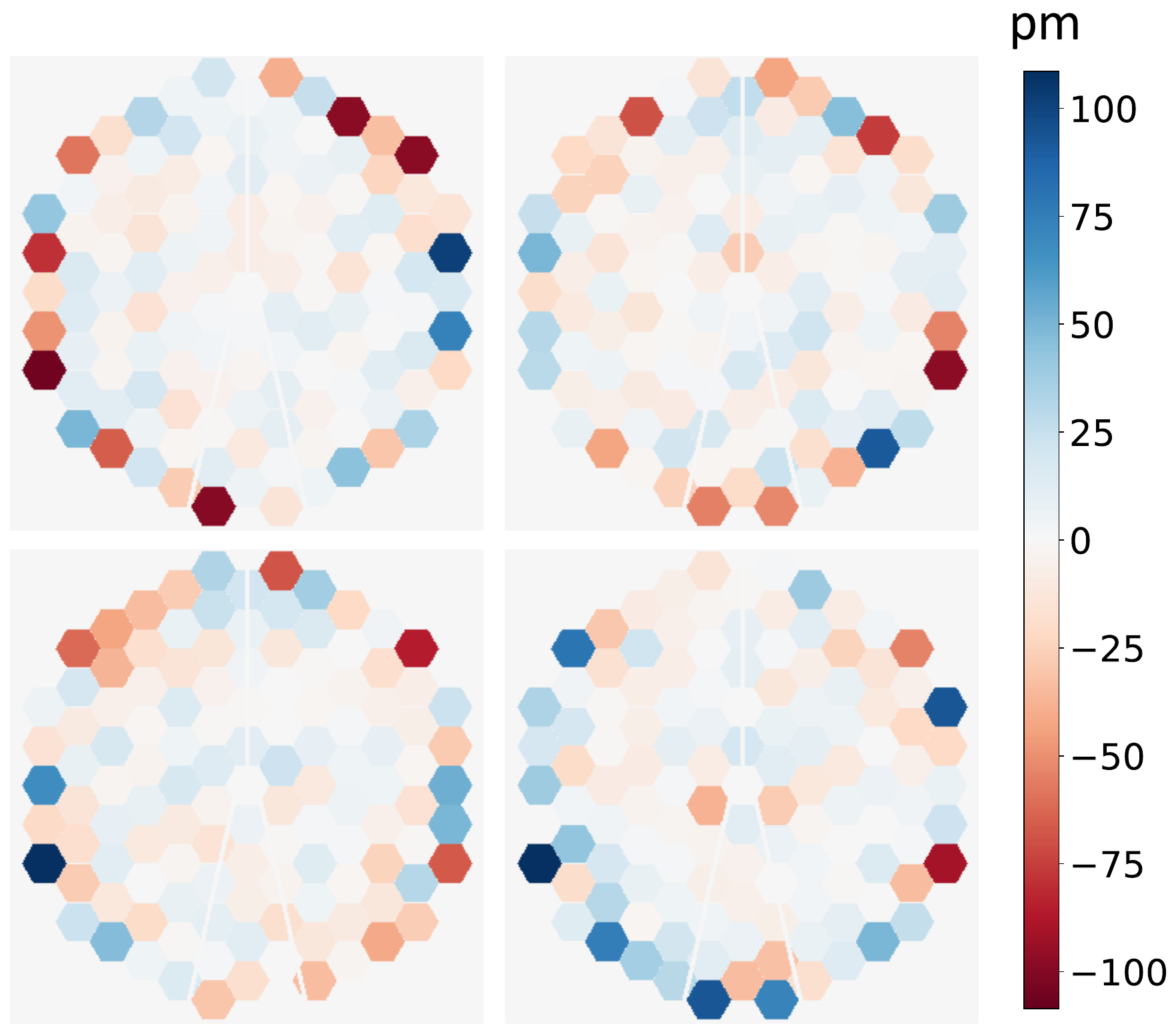}
   \end{tabular}
   \end{center}
   \caption[segment-map-small] 
   {\label{fig:random-maps} 
    Four random segment-based WFE maps drawn from a zero-mean normal distribution and the per-segment standard deviations from the left prescription map in Fig.~\ref{fig:segment-maps}, for the narrow-angle APLC design and a target contrast of $10^{-10}$. After each random map is created, we propagate it through the end-to-end simulator and record its average contrast to build the left MC simulation in Fig.~\ref{fig:histograms_three_contrasts}.}
   \end{figure}

\begin{figure*}
   \begin{center}
   \begin{tabular}{@{}c@{}}
   \includegraphics[width=\linewidth]{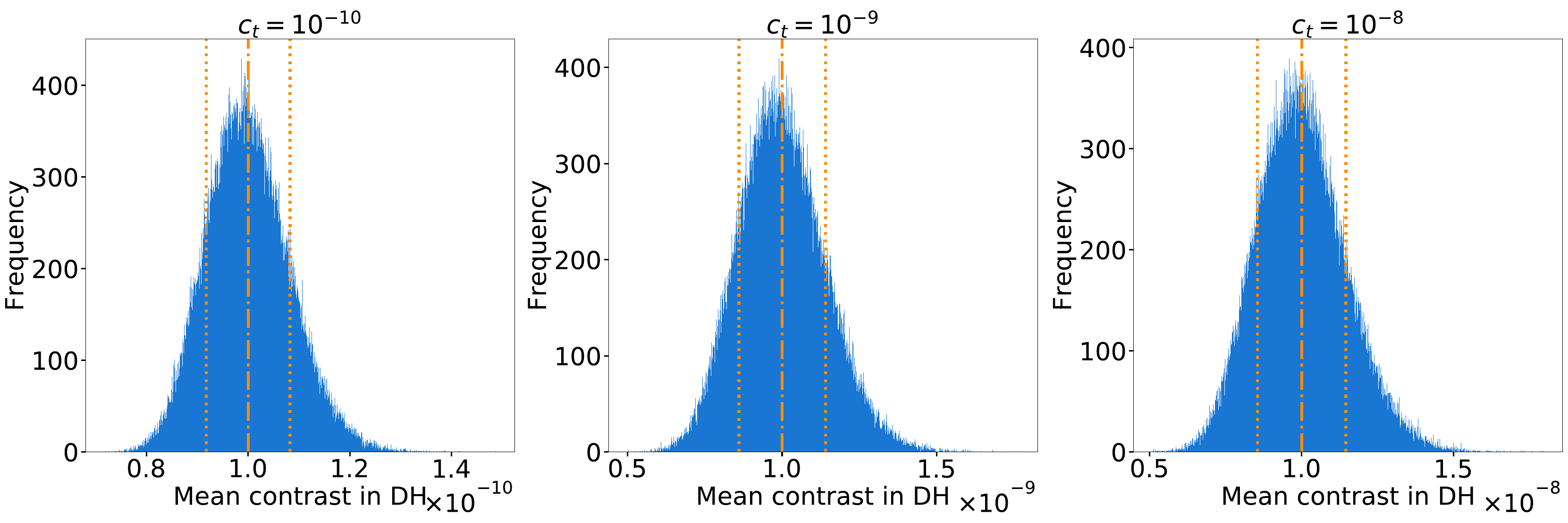}
   \end{tabular}
   \end{center}
   \caption[segment-map-small] 
   {\label{fig:histograms_three_contrasts} 
    Validation of the independent segment tolerancing with E2E Monte-Carlo simulations, for different target contrasts, using the narrow-angle APLC design. Each segment $k$ in one of the 100,000 WFE realizations is drawn from a zero-mean normal distribution with standard deviation $\mu_k$. The dashed-dotted lines mark the target contrast of each case, which are successfully recovered by the mean values of the histograms, in accordance with their analytical calculation in Eq.~\ref{eq:avg-contrast-from-trace}. The dotted lines mark the 1-sigma confidence limits of this contrast distribution, which are 8.3$\times10^{-12}$, 1.4$\times10^{-10}$ and 1.5$\times10^{-9}$ for the three target contrasts $10^{-10}$, $10^{-9}$ and $10^{-8}$ respectively, and they accord with the numbers calculated by Eq.~\ref{eq:var-of-c}.}
   \end{figure*}

It is important to note that these requirement maps do not represent the WFE over the segmented pupil, but instead show the standard deviations on the tolerable WFE for each segment in order to retrieve, statistically, the desired mean target contrast. In this sense, the maps in Fig.~\ref{fig:segment-maps} are a prescription for the drawing of random segment WFE realizations like the examples shown in Fig.~\ref{fig:random-maps}. These random maps are then propagated with the E2E simulator and their average contrast values build the MC histograms in Fig.~\ref{fig:histograms_three_contrasts}.
One big takeaway point from Fig.~\ref{fig:segment-maps} is that the segment requirements are not uniform across the pupil, but clearly follow the apodization of the coronagraph mask. The PASTIS matrix holds knowledge of the optical effect of not only the segments but also the coronagraph instrument on the final contrast, so by including that knowledge into the derivation of the segment constraints we obtain a requirement map optimized for that particular instrument. Moreover, we can observe a direct trade-off between the coronagraph apodization and the per-segment requirements - the more aggressive the apodization and the lower the throughput, the more we can relax the requirements on the more concealed segments within one coronagraph. However, more aggressive apodization usually comes with smaller FPM coronagraphs that filter low-order modes less, which will lead to more stringent overall requirements. This leads to a direct trade-off between FPM size, throughput and WFE requirements (see also Sec.~\ref{sec:DISCUSSION}).

These requirement maps can be calculated for any target contrast in the range of validity of the PASTIS model, which we discussed in Sec.~\ref{subsec:validating-pastis}.  We can verify Eq.~\ref{eq:single-mus} by running MC simulations with the E2E simulator across a grid of different coronagraph instruments and target contrasts. Using a range of target contrasts $c_t = 10^{-10}, 10^{-9}$, and $10^{-8}$ on the narrow-angle baseline LUVOIR APLC design, we first calculate the segment constraints (the requirement maps for $ 10^{-9}$ and $ 10^{-8}$ are not shown here, but they show the same spatial distribution over the segments as in Fig.~\ref{fig:segment-maps}, only different by a proportionality factor). We draw the WFE amplitude for each individual segment from a zero-mean normal distribution and its standard deviation $\mu_k$:
    \begin{equation}
    \mathbf{a} = (\mathcal{N}(0,\mu_1),\ \mathcal{N}(0,\mu_2),\ \ldots,\ \mathcal{N}(0,\mu_k)).
    \label{eq:a-normal-distribution}
    \end{equation}
We use these random aberration amplitudes on all segments to compose a WFE map on the segmented pupil and then propagate this WFE map through the E2E simulator to measure the resulting spatial average contrast in the dark hole. Doing this 100,000 times for each target contrast case, we obtain the histograms shown in Fig.~\ref{fig:histograms_three_contrasts}. The mean of the resulting MC simulations clearly recovers the target contrast for which the segment requirements have been calculated, which is indicated by the dashed-dotted line. Both these mean values, as well as the standard deviations, indicated with the dotted lines in Fig.~\ref{fig:histograms_three_contrasts}, agree with the theoretical values calculated analytically from the segment covariance matrix (Eq.~\ref{eq:avg-contrast-from-trace} and Eq.~\ref{eq:var-of-c}). Also, we have verified the correct recovery of the same range of target contrasts by means of MC simulations for the other two APLC designs shown in Fig.~\ref{fig:segment-maps} (resulting histograms not shown in this paper). 

\subsection{Modal analysis of the segment-based requirements}
\label{subsec:optimized-error-budget}

\begin{figure}
   \begin{center}
   \begin{tabular}{c}
   \includegraphics[width = \textwidth]{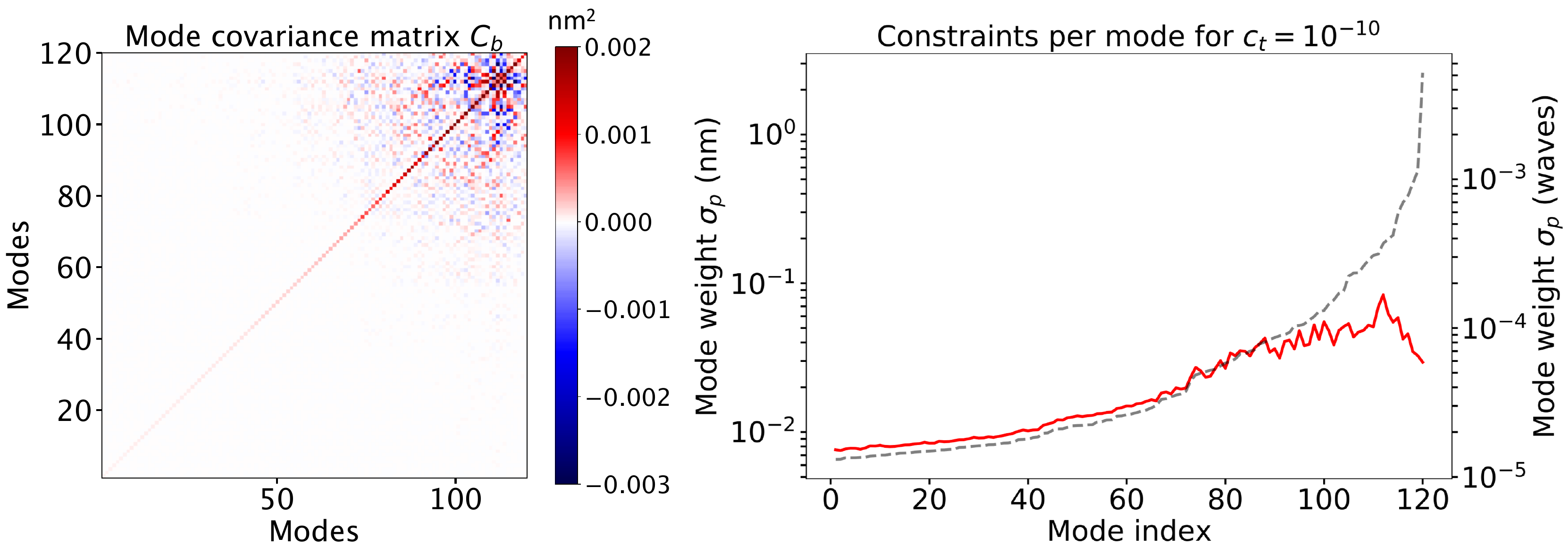}
   \end{tabular}
   \end{center}
   \caption[optimized-sigmas-small] 
   {\label{fig:optimized-sigmas-small} 
    \textit{Left:} Covariance matrix $C_b$, calculated from the diagonal covariance matrix in segment-space, $C_a$, with $C_b = U^{T} C_a U$. Although there are clearly some correlations present between the high-index PASTIS modes in the top right corner (low spatial frequencies), this does not matter as long as we are in the PASTIS segment basis, where the PASTIS matrix is diagonal. When this is the case, the mean contrast only depends on the diagonal elements of $C_b$ (Eq.~\ref{eq:statistical-sigmas}). \textit{Right:} PASTIS mode amplitudes for the case of independent segments in WFE RMS (solid red). They are extracted from the mode covariance matrix $C_b$, after constructing an error budget assuming independent segments that contribute equally to the total contrast, at 500~nm. Overlapping (dashed grey), we can see the mode weights from the uniform contrast allocation to all PASTIS modes from Fig.~\ref{fig:sigmas-flat-error-budget}. We can clearly see how compared to that flat allocation, the independent-segment error budget increases the tolerances of low-index modes (left) that have less segment correlation, and dampens the tolerances of high-index modes (right) that are highly correlated, low-spatial frequency modes.}
   \end{figure}

\begin{figure}
   \begin{center}
   \begin{tabular}{c}
   \includegraphics[width = 10cm]{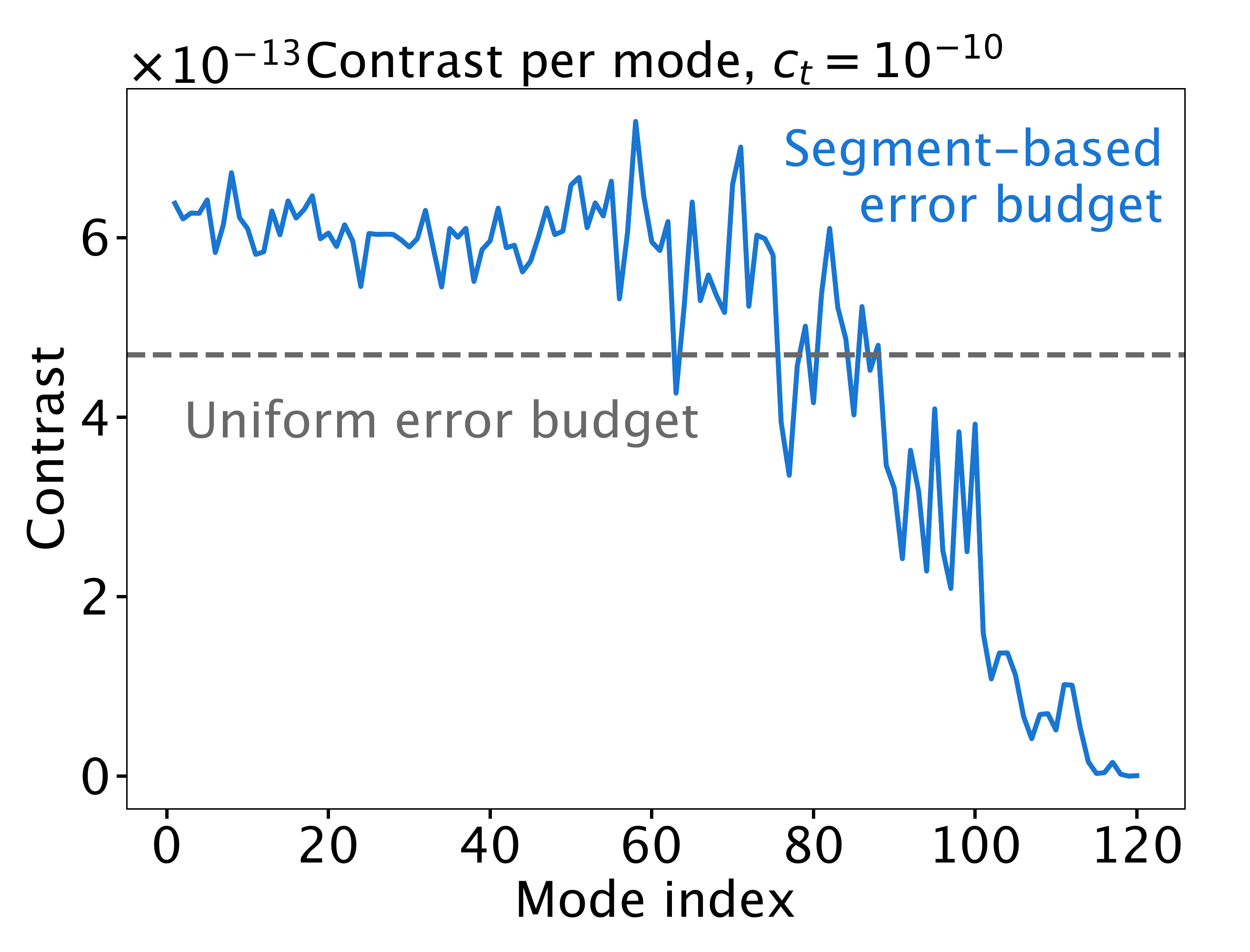}
   \end{tabular}
   \end{center}
   \caption[optimized-contrast-per-mode] 
   {\label{fig:optimized-contrast-per-mode} 
    Contrast per individual PASTIS mode when derived from the error budget in which all segments contribute independently and equally to the final contrast (solid blue). High-index modes to the right, which correspond to low-spatial frequencies and therefore highly correlated segments (see Fig.~\ref{fig:postage-stamp-modes}), are highly attenuated and contribute negligible amounts to the contrast. The uniform contrast allocation across all modes at $(c_t - c_0) / n_{seg}$ is indicated with the dashed grey line (the contrast floor has been removed in both curves).}
   \end{figure}

The segment requirement maps were obtained assuming a uniform contrast allocation across all segments (Eq.~\ref{eq:segment-requirement}). We also assumed statistically independent segments, so that their correlation matrix $C_a$ was diagonal. Here, we further explore this uniform error budget in the segment basis by analyzing the corresponding distribution of proper system modes of the optical system, the PASTIS modes. Using the transformation matrix $U$ from the eigendecomposition of the PASTIS matrix $M$, we can calculate the corresponding covariance matrix in the PASTIS mode basis with $C_b = U^{T} C_a U$. Given this linear transformation, if the covariance matrix is diagonal in one space, we do not expect it to be diagonal in the other space. The matrix $C_b$, obtained from the diagonal segment covariance matrix $C_a$ assembled from the requirement map, is illustrated on the left hand side in Fig.~\ref{fig:optimized-sigmas-small}. This figure also compares the extracted standard deviations for PASTIS modes along the diagonal of $C_b$, with the PASTIS mode weights previously calculated in Sec.~\ref{sec:model-inversion}, on the right hand side. Although $C_b$ is not diagonal, this is a legitimate comparison. The PASTIS matrix $M$ is always diagonal in its own eigenbasis, expressed as matrix $D$ in Sec.~\ref{subsec:pastis-eigendecomposition}. This is why the average contrast expression from the statistical mode weights in Eq.~\ref{eq:statistical-sigmas} only requires the diagonal elements of the covariance matrix $C_b$, no matter whether it is diagonal or not, i.e. whether the mode weights show some correlations or not. The difference with respect to the mode weights $\widetilde{b_p}$ (Eq.~\ref{eq:calc-sigma-uniform}) obtained under the assumption of a uniform contrast allocation per mode (Fig.~\ref{fig:sigmas-flat-error-budget}) is notable: the mode weights of low mode index have increased tolerances, which is very interesting from a system design point of view, while large index modes (above $\sim$90) are strongly attenuated in the PASTIS mode basis error budget for independent segments. This is also clearly visible in Fig.~\ref{fig:optimized-contrast-per-mode} where the contrast contribution per mode is relatively flat at a low mode index, but drops to negligible contributions at high-index modes. For the case of the flat contrast allocation across modes, this figure shows a flat line at $(c_t - c_0) / n_{seg}$ for comparison (dashed line).
This effect can be well understood by looking back at Fig.~\ref{fig:postage-stamp-modes}, where high-index modes appear to be very similar to low-order Zernike modes, therefore having highly correlated segments, and the low-index modes appear as high-spatial frequencies, i.e. with more uncorrelated segments. Therefore, the construction of a segment-level error budget for uncorrelated segments creates a modal distribution with extremely low weights on the PASTIS modes that have highly correlated segments (high-index modes), as seen in Fig.~\ref{fig:optimized-sigmas-small}. Also, since the mode contrast contribution is directly related to the mode weight (Eq.~\ref{eq:calc-sigma}), the same effect is visible in the allocated contrast per mode (Fig.~\ref{fig:optimized-contrast-per-mode}).
  
\begin{figure}
   \begin{center}
   \begin{tabular}{c}
   \includegraphics[width = 10cm]{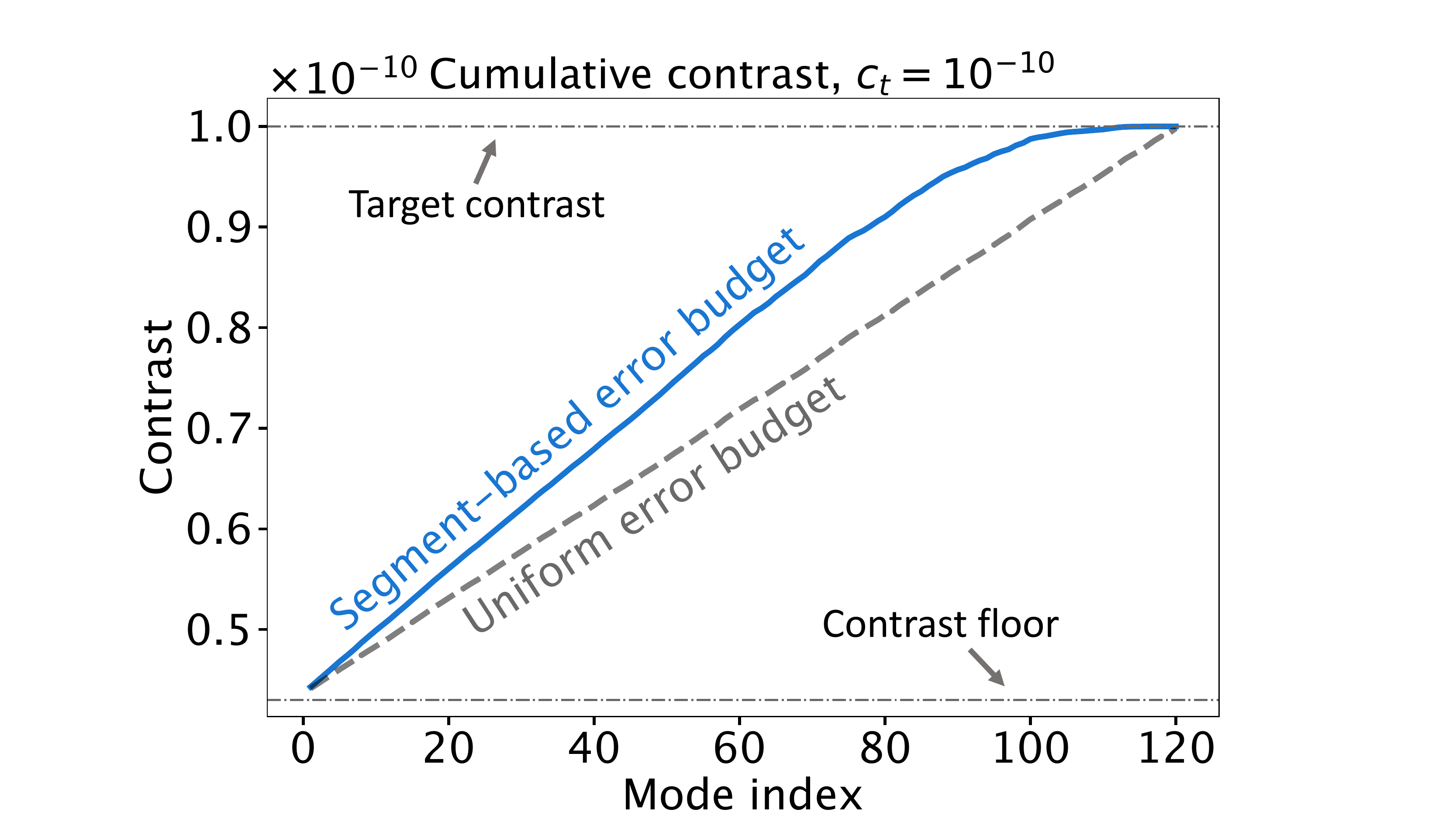}
   \end{tabular}
   \end{center}
   \caption[optimized-cumulative-contrast] 
   {\label{fig:optimized-cumulative-contrast} 
    Cumulative contrast of the PASTIS modes for the error budget in which all segments contribute independently and equally to the final contrast (solid blue), compared to the case in Fig.~\ref{fig:cumulative-contrast} where all modes contribute the same contrast (dashed grey). The high-index modes have negligible contrast impact (see also Fig.~\ref{fig:optimized-contrast-per-mode}) as they correspond to low-spatial frequency, highly correlated segments. This plot also confirms the assumption that the mode covariance matrix $C_b$ is nearly diagonal.}
   \end{figure}

To illustrate this further, we calculate a cumulative contrast plot similar to Fig.~\ref{fig:cumulative-contrast}, which was initially obtained for a uniform contrast allocation per mode. The new result is shown in Fig.~\ref{fig:optimized-cumulative-contrast}, where we can see that it is no longer linear, i.e. the modes no longer contribute equally to the total mean contrast. The slope of the blue curve is indicative of the allocated tolerances for each mode: the low-spatial frequency PASTIS modes on the right hand side now contribute significantly less to the final contrast, while the first $\sim80$ modes contribute more, while still resulting in the exact cumulative target contrast. This is consistent with the behavior discussed in Figs.~\ref{fig:optimized-sigmas-small} and \ref{fig:optimized-contrast-per-mode}.

\section{DISCUSSION}
\label{sec:DISCUSSION}

The results we obtain for the three LUVOIR APLC designs in Sec.~\ref{sec:APPLICATION-TO-LUVOIR}, under the assumption of statistically independent segments, span more than one order of magnitude from the most constrained segment on the small APLC to the most relaxed segment on the large APLC (7~pm to 181~pm). These results show not only a dependency on the coronagraph design, but also a wide range of segment requirements within the pupil for one single coronagraph. The segment with the most stringent requirement can tolerate a standard deviation of 7~pm local piston error on the narrow-angle APLC for a target contrast of $10^{-10}$, at a wavelength of 500~nm, which is comparable to previous results for segment-based piston errors on LUVOIR-A (10~pm at a wavelength of 575~nm\cite{Juanola-Parramon2019}), while studies on other apertures and with other segment numbers quote similar numbers\cite{Stahl2015, Nemati2017}. 
However, we show that segments in other parts of the pupil have a much higher local piston tolerance, a standard deviation up to 116~pm on the small APLC design, which suggests that not all segments need to be held to the same tolerance level. Instead, we can relax the segment-level requirements on those segments that do not influence the average DH contrast as much, while still obtaining the same statistical mean contrast. Local relaxation of the wavefront error limits on certain parts of the pupil can be exploited for example for the backplane mechanical design and observatory-level control strategy.

In particular, the tolerances will also depend on the total number of segments in the pupil. While this has not been studied systematically in this paper, PASTIS can enable such work. For example, with fewer segments in the aperture, the spatial frequencies corresponding to segment misalignments will be lower. Therefore, WFE from these misalignments will be more filtered by the coronagraph, which might lead to increased tolerances. Inversely, with more segments in the pupil, the highest spatial frequencies from segment misalignments will not be filtered by the FPM and might thus result in lower tolerances, as is already the case for the most sensitive PASTIS modes. Wavefront aberrations from a mirror with  fewer segments will be typically more filtered by the coronagraph. However, more and smaller segments will introduce high spatial frequencies that will diffract light into the image beyond the outer working angle, decreasing the impact of each segment misalignment on the DH. Overall, the number of segments in the pupil will raise competing effects that will influence the overall tolerances, and will be highly influenced by the type of coronagraph\cite{Laginja2020SPIE}.

In Sec.~\ref{subsec:MC-on-LUVOIR-segments}, we briefly mentioned the observed trade-off between coronagraph throughput, FPM size and per-segment requirements. For a given coronagraph design, we observe relaxed requirements for those segments in the pupil that are more concealed by the apodizer (i.e., more black in the apodizer image). The overall tolerance over the entire pupil also increases with the size of the FPM.
The larger the FPM, the higher the rejection in particular of low-order spatial modes, which correspond to high mode indices where the modes are similar to low-order, Zernike-like global modes. This results in higher mode weights, which becomes obvious in Fig.~\ref{fig:sigmas-flat-error-budget}, where these high-index modes on the right side show higher WFE tolerances. When moving to the independent-segment error budget in Fig.~\ref{fig:optimized-sigmas-small}, this effect becomes less obvious as the low-order mode tolerances get dampened due to our assumption of uncorrelated segments contributing equally to the mean contrast. The projection of these uncorrelated segments onto the mode basis favors high-spatial frequency modes. This leaves the low-order modes statistically weak, as they would otherwise contribute to inter-segment correlation.
However, this is only true for these fully uncorrelated segment-level WFE contributions. This uncorrelated error budget is over-constraining the low-order (high-index) modes (see Fig.~\ref{fig:optimized-sigmas-small}), therefore not taking advantage of the coronagraphic rejection of these spatially correlated modes. A complete error budget will need to allocate contrast contributions separately between the correlated and the uncorrelated components of the WFE. The final result will therefore have a modal weight distribution in-between the solid (fully uncorrelated) and dashed lines (uniform contrast across modes) in Fig.~\ref{fig:optimized-sigmas-small}.

Additionally, large-FPM APLCs have higher throughput apodizers. This means that their larger coronagraphic rejection (associated with the larger FPM) contributes more to the WFE tolerance relaxation than the apodizer throughput itself. Therefore, a true optimization of the WFE tolerances will be a trade-off between the FPM size and the fraction of apodization in the pupil. Further, this tolerancing work introduces new design considerations for high-contrast instruments, which is the optimization of the coronagraphic component with respect to segment phasing tolerances. Such an optimization will aim to release the segment tolerances while keeping a reasonable contrast goal, with the ultimate goal to maximize exoplanet yield, which should be explored in future work.

In a realistic telescope of course, the segments are typically correlated due to the deformations of the backplane structure. The development of technologies that support increased wavefront stability of segmented telescopes is actively being worked on today\cite{Coyle2019}. This includes precise methods for thermo-mechanical modelling, and measuring of such deformation effects on the segmented primary mirror. The underlying segment correlations can be provided either as a segment covariance matrix, or in the form of thermo-mechanical system eigenmodes, in which case we can directly use Eq.~\ref{eq:avg-contrast-diagonal} for a tolerancing analysis, after expressing the PASTIS matrix in this new basis.

While local piston errors have been shown to have the largest impact on the contrast\cite{Nemati2017, Juanola-Parramon2019}, we can generate a PASTIS matrix for other local Zernike modes as well (e.g. tip/tilt, focus, astigmatism). The feasibility of this has already been shown in the analytical approach\cite{Leboulleux2018jatis} and should hence be regarded as a mere functional addition. Moreover, if we have knowledge of telescope-design and hardware dependent local aberration modes (e.g. from effects such as adhesive shrinkage, bulk temperature, coatings, or gravity), these can be used as well to build the PASTIS matrix and derive their corresponding segment-level requirements. Instead of evaluating the tolerance levels mode by mode, we can also calculate a multi-mode PASTIS matrix that incorporates combinations of local modes (e.g. piston--tip--tilt, or combinations of custom modes) in order to derive segment tolerances that will take into account that more than one distinct local aberration mode is contributing to the overall WFE.

Another application of extended PASTIS matrices is the generalization to high-spatial frequency effects (e.g. from polishing). Instead of building a PASTIS matrix with pair-wise Zernike aberrations of segments, we can use sine waves locally on the segments in lieu of Zernikes. Each spatial frequency and orientation would then be a new local mode, and we can use many of them to make a multi-mode PASTIS matrix in the same way as with any other local modes. While a generalization to a continuous distribution of frequencies to build tolerances in terms of a Power Spectral Density (PSD) might be possible, it is beyond the scope of the present discussion. Nevertheless, the generalization to a few sine-wave frequencies (e.g with their corresponding speckles localized at the inner and outer working angle, or in the middle of the DH) is a direct and straightforward extension of the present illustration and would provide meaningful input for tolerancing purposes of polishing errors.

The presented tolerancing model provides WFE limits on the segments, but it does not define how these limits are to be maintained. Relying purely on the mechanical stability of the telescope will not be enough to stay within these requirements and therefore an active optics system will be needed to measure WFE deviations and compensate for the residuals. Such an active optics system will include WFS\&C as well as signal-to-noise considerations, as the wavefront sensor needs enough photons to provide an accurate wavefront estimate and correction\cite{Pueyo2019AAS}.

Finally, the application of PASTIS to ground-based observatories is possible, but will have to take additional effects in account. While future large segmented telescopes (TMT, E-ELT, GMT) will reach contrast levels of $10^{-7}$ to $10^{-8}$ within the next decade\cite{Kasper2008}, which is sufficient to enter the high-contrast regime that the PASTIS model can be applied to, future work will have to include the effects of residual turbulence in order to truthfully represent coronagraphic observations on those observatories. As a first common approximation for this purpose, the coronagraphic PSF can be expressed as the sum of a static and a dynamic contribution\cite{Ygouf2013}, so the PASTIS analysis can be used for a characterization of the static part.

\section{CONCLUSIONS}
\label{sec:CONCLUSION}

The goal of PASTIS, as established by Leboulleux et al.\cite{Leboulleux2018jatis}, is an analytical propagation model to calculate the average dark hole contrast in a coronagraphic system, in the presence of segment-level aberrations. This is achieved with a closed-form expression in Eq.~\ref{eq:pastis-equation} that depends exclusively on the PASTIS matrix $M$, acting on the aberration amplitudes on all segments. In this paper, we extended the calculation of the matrix $M$ to a semi-analytical approach, where the optical propagation of segment aberrations is performed numerically before assembling the $M$ matrix analytically (Eqs.~\ref{eq:diagonal-elements} and \ref{eq:off-diagonal-elements}). This makes the model more accurate as it includes all details of the optical system as provided with the end-to-end simulator. We also show that the model holds even for a non-symmetrical DH. The semi-analytical PASTIS approach is therefore a flexible tolerancing tool that can be adapted readily to any telescope geometry or coronagraph, as shown in Sec.~\ref{sec:APPLICATION-TO-LUVOIR}. It can be used to study trade-offs between coronagraph designs that will provide certain tolerance distributions over the segments on the primary mirror, and telescope-level engineering constraints implemented in other parts of the observatory.

We used the model to derive analytical expressions for the statistical mean and variance of the average DH contrast (Eqs.~\ref{eq:avg-contrast-from-trace} and \ref{eq:var-of-c}). This opens the possibility for WFE tolerancing of a segmented observatory. In addition to the optical properties modeled by the PASTIS matrix $M$, these expressions involve the segment-level covariance matrix $C_a$ that describes the thermo-mechanical properties of the telescope. Indeed, deformations of the backplane structure typically lead to correlated segment poses (e.g. backplane flapping modes around the folding motion of the primary mirror).  It is this combination of the optical with the thermo-mechanical characteristics that lays the foundation to a complete and analytical method for the tolerancing of segmented aberrations.

The key to calculate WFE requirements with this framework is to find the diagonal basis of either of the two matrices, $M$ or $C_a$. The PASTIS matrix $M$ can easily be diagonalized by means of an eigendecomposition, in which case the tolerancing can be performed on the PASTIS eigenmodes. They form an orthonormal set of modes, representing the proper optical system modes of the given observatory and coronagraphic instrument, and allow to analyze its fundamental limitations in terms of WFE propagation in segmented aperture coronagraphy. These eigenmodes contribute additively to the image plane average contrast according to their mode-level tolerances, which correspond to the standard deviations per mode $\sigma_p$, associated with the statistical mean of the average DH contrast.

Additionally, if we have information on the thermo-mechanical behavior of the instrument or of the whole observatory embodied in the segment aberration covariance matrix (e.g., though finite-element simulations), we can choose to work in the eigenbasis of these thermo-mechanical perturbations, a.k.a. the Karhunen-Lo\`eve basis, to perform the tolerancing.

This allows us to put requirements on structural deformations that impact the segmented primary mirror (e.g. backplane and mirror support structures) and can be modelled as rigid-body motions at the segment level. A better approximation of these thermo-mechanical modes will be obtained with a multi-mode PASTIS matrix by combining multiple local Zernikes and/or ad-hoc local aberration modes, as described in Sec.~\ref{sec:DISCUSSION}.


In the simplified case of independent segments, the covariance matrix $C_a$ contains only diagonal elements, representing the individual segment WFE variances $\mu_k^2$. We have built a segment-based error budget by allocating equal contribution to contrast from all segments. This allowed us to calculate WFE requirements for all segments individually, building segment requirement maps as we showed in Sec.~\ref{sec:APPLICATION-TO-LUVOIR} for the three APLC designs of the LUVOIR-A telescope. The advantage of this method is two-fold: Firstly, rather than calculating WFE tolerances globally over the entire pupil, we can obtain a WFE standard deviation per segment, which can locally lead to a relaxation in requirements. Secondly, we do not need to perform full Monte-Carlo simulations that evaluate different realizations of wavefront error maps for that purpose, instead, we can calculate these requirements analytically in one single step.

The analysis presented in this paper is statistical but static, i.e., without temporal evolution; the extension to dynamical drift rates depends on the observing scenario and wavefront control strategy, which will put the PASTIS propagation model on different time scales\cite{Coyle2019, Pueyo2019}. Future work will address such dynamic analysis methods for continuous wavefront sensing and control cases on ultra-stable telescopes.

\acknowledgments 
This work was co-authored by employees of BALL AEROSPACE as part of the the Ultra-Stable Telescope Research and Analysis (ULTRA) Program under Contract No. 80MSFC20C0018 with the National Aeronautics and Space Administration (PI: L. Coyle), and by STScI employees under corresponding subcontracts No.18KMB00077 and No.19KMB00102 with Ball Aerospace (PI: R. Soummer, Sci-PI: L. Pueyo). This work was also co-authored by employees of the French National Aerospace Research Center ONERA (Office National d'\'{E}tudes et de Recherches A\'{e}rospatiales), and benefited from the support of the WOLF project ANR-18-CE31-0018 of the French National Research Agency (ANR), as well as the internal research project VASCO. This work was also supported in part by the National Aeronautics and Space Administration under Grant 80NSSC19K0120 issued through the Strategic Astrophysics Technology/Technology Demonstration for Exoplanet Missions Program (SAT-TDEM; PI: R. Soummer), and by the Segmented-aperture Coronagraph Design and Analysis funded by ExEP, under JPL subcontract No.1539872, and by the STScI Director's Discretionary Research Funds.  The authors acknowledge the contributions of Kathryn St.Laurent, James Noss and Emiel Por to the design and optimization of the LUVOIR-A apodizers. I. Laginja is thankful to Th\'{e}o Jolivet and Anand Sivaramakrishnan for helpful discussions.
The United States Government retains and the publisher, by accepting the article for publication, acknowledges that the United States Government retains a non-exclusive, paid-up, irrevocable, worldwide license to reproduce, prepare derivative works, distribute copies to the public, and perform publicly and display publicly, or allow others to do so, for United States Government purposes. All other rights are reserved by the copyright owner. 

This research was developed in Python\footnote{\url{https://www.python.org}}, an open source programming language, and made use of the Numpy\cite{Oliphant2006numpy, vanderWalt2011numpy}, Matplotlib\cite{Hunter2007matplotlib, Caswell2019matplotlib-zenodo}, Astropy\cite{AstropyCollaboration2013, AstropyCollaboration2018, Astropy2018zenodo} and HCIPy\cite{Por2018hcipy} packages. The software for all analysis was published in PASTIS\cite{Laginja2020pastis}, a modular, open-source Python package for segment-level error budgeting of segmented telescopes.

\appendix
\section{LUVOIR-A end-to-end simulator}
\label{appendix-simulator}
The LUVOIR-A end-to-end simulator used in this paper is fully written in Python, and uses the HCIPy\cite{Por2018hcipy} package for optical propagations and the Lyot coronagraph implementation. The simulator inherits from the HCIPy SegmentedMirror class that provides the capability to individually actuate the segments in a telescope pupil. The core of the simulator is an abstract class that combines this segmented actuation with an APLC, which gets inherited by the the main LUVOIR-A simulator that defines the specific aperture geometry, apodizers, FPMs and Lyot stops of the LUVOIR-A APLCs. The source files (aperture, apodizers, Lyot stop) of this APLC suite are credited to the Segmented-aperture Coronagraph  Design  and  Analysis (SCDA) study, and the user can switch between the three designs with a single keyword parameter. The simulator calculates the electric field at each pupil and focal plane of the optical system, including the final image plane of the coronagraph. These electric fields are then returned by the simulator.

Correct optical propagation of various coronagraphs, including the APLC, has been implemented by various other simulators, as have been segmented mirrors. The problem that this particular simulator solves and makes available is that of proper sampling of the pupil for the segmented actuation with respect to the pupil plane coronagraph optics: the apodizer and the Lyot stop. It is very important that the pixels of the individual segments overlap perfectly with the apodizer, otherwise the coronagraph will not perform well. Both the aperture as well as the apodizer design and Lyot stop are input parameters to the simulator, hence the correct pixel-to-pixel mapping between the aperture and the apodizer are guaranteed.

Currently, the segmented mirror control includes local piston, tip and tilt, but can be easily extended to other modes and modal bases by implementing newer versions of HCIPy, which is ongoing refactoring. The simulator version used for this paper is published within the PASTIS Python package\cite{Laginja2020pastis} and available on GitHub.

\bibliography{2020jatis}
\bibliographystyle{spiebib}

\end{document}